\documentclass[hidelinks]{article}
\usepackage[margin=1in]{geometry}
\usepackage[dvips]{graphicx}
\usepackage{dcolumn,amsmath,amssymb}
\usepackage{bm}
\usepackage[caption=false]{subfig}
\usepackage{hyperref}
\usepackage{slashed,physics}
\usepackage{amsfonts,multirow,array,multirow,booktabs,mathtools}
\usepackage{xcolor,appendix,float}
\newcommand{\Lagr}{\mathcal{L}}

\graphicspath{ {IMAGES/} }
\begin{document}
\title{Analysis of 2S singly heavy baryons in HQET}
\author{K. K. Vishwakarma\thanks{vish.kumar.kundan@gmail.com}, Alka Upadhyay}
\date{%
    School of Physics and Materials Science, Thapar Institute of Engineering and Technology, Patiala-147004, INDIA}
\maketitle
\begin{abstract}We have employed HQET to determine the masses of radially excited ($n=2$) S-wave charm and bottom baryons. The HQET Lagrangian containing the non-perturbative parameters is shown with heavy baryon fields. The non-perturbative parameters, couplings, and decay widths are also studied for the S-wave singly heavy baryons. The HQET parameters $\overline{\Lambda}$, $\lambda_1$ and $\lambda_2$  are calculated for the ground state ($n=1$) using the masses of S-wave baryons. The mass term ratios of $n=1$ and $n=2$ mesons and baryons containing parameters $\overline{\Lambda}$ and $\lambda_1$ are studied by varying the bottom quark mass. This analysis shows that heavy quark behaves the same inside mesons and baryons in both 1S and 2S states. The HQET symmetry of $\overline{\Lambda}$ is used to find the parameters and masses for $n=2$ S-wave baryons. The variation of mass of 2S baryons with the non-perturbative parameters $\lambda_1$ and $\lambda_2$ is discussed. The Regge trajectories are also plotted in the $(n,M^2)$ plane using masses of $n=1$ and 2 charm and bottom baryons. The Regge trajectories are parallel and equidistant lines in the $(n,M^2)$ plane. We have also studied the strong decays of charm and bottom baryons for both $n=1$ and $n=2$ states. We have estimated the coupling constants $g_1=0.913^{+0.010}_{-0.017}$ and $g_2=0.559^{+0.006}_{-0.010}$ for $n=1$, and $\frac{\widetilde{g}_1}{\widetilde{g}_2}=1.52$ for $n=2$. We have also shown the semi-electronic decays rates of charm baryons in the spectator heavy quark approximation for $1S\rightarrow 1S$, $2S\rightarrow 1S$ and $2S\rightarrow 2S$ transitions. The decay rates for $1S\rightarrow 1S$ transitions are of same order as $2S\rightarrow 1S$ transitions. This analysis gives a good agreement with available theoretical and experimental data.
\end{abstract}
\section{Introduction}
Heavy light systems containing a single heavy quark are an active area of research due to continuous experimental observations. The radial excitation of these heavy baryons lies in the same mass regions as many of the recently observed baryons.
Despite new observations \cite{LHC2019PRL123,LHCb2019PRL122,LHCb2020PRL124,Belle2021PRD} of baryons, the spectrum for radially excited charm and bottom baryons is not much explored. In PDG \cite{PDG2021},  we find only few radially excited baryons candidates. The state $\Lambda_c(2765)/\Sigma_c(2765)$ was observed by CLEO collaboration \cite{CLEO2001} in 2001, and its isospin was confirmed to be zero by Belle collaboration \cite{Belle2019conference} in 2019. This zero isospin established the state as a $\Lambda_c$ resonance. The $\Lambda_c(2765)$ is predicted \cite{Chen2015,Chengcirca2021,Galkin2020} as the first radial excitation, $\Lambda_c(2S)$. Belle collaboration in 2006 observed two peaks $\Xi_c(2970)$ and $\Xi_c(3077)^+$ in decay channel $\Lambda_c^+K^-\pi^+$ \cite{Belle2006}. The same state $\Xi_c(2970)^+$ was identified in decay modes $\Xi_c(2645)\pi$ \cite{Belle2008}, $\Xi_c^{'}\pi$ \cite{belle2016} by Belle collaboration and $\Sigma_c(2455)K$ by BaBar collaboration \cite{Babar2008}. 
In 2021, Belle collaboration \cite{Belle2021Xi} identified the spin parity of the $\Xi_c(2970)$ to be $\frac{1}{2}^+$ using 980 fb$^{-1}$ data sample collected by the Belle detector at the KEKB asymmetric-energy $e^+e^-$ collider with the light degrees of freedom $s_l=0$. The assignment of $\Xi_c(2970)$ was theoretically studied by \cite{Chen2015,Chengcirca2021,Galkin2020,Cheng2007Xi2970,ebert2011,chengHHChPT2015,Wang2020} and supported by Belle \cite{Belle2021Xi} to be $\Xi_c(2S)$ . 
In 2017, LHCb \cite{LHCbomegac2017} observed five new narrow excited $\Omega_c^0$ states in the $\Xi_c^+K^-$ mass spectrum with the sample of pp collision data corresponding to an integrated luminosity of 3.3 fb$^{-1}$, collected by the LHCb experiment. The states were $\Omega_c(3000)^0$, $\Omega_c(3055)^0$, $\Omega_c(3066)^0$, $\Omega_c(3090)^0$ and $\Omega_c(3119)^0$. LHCb Collaboration in 2021 \cite{LHCbOmega2021}, suggested assignments to the four observed resonances $\Omega_c(3000)^0$, $\Omega_c(3055)^0$, $\Omega_c(3066)^0$, $\Omega_c(3090)^0$ to be the $\lambda-$mode excitation with $J^P=\frac{1}{2}^-$, $\frac{3}{2}^-$, $\frac{3}{2}^-$ and $\frac{5}{2}^-$ respectively. 
The absence of $\Omega_c(3119)^0$ indicated that it may be the first radial excitation $\Omega_c(2S)$ with spin $\frac{1}{2}^+$ or $\frac{3}{2}^+$ \cite{Galkin2020,Wang2017} or the $\rho-$mode excitation of $P$-wave \cite{Chengcirca2021}.

In 2020, LHCb collaboration \cite{LHCb2020lambda} observed a new baryon state in the $\Lambda_b^0\pi^+\pi^-$ mass spectrum with mass $m_{\Lambda^{**0}_b}=6072.3\pm2.9\pm0.6\pm0.2$ MeV and natural width $\Gamma=72\pm11\pm2$ MeV.
The LHCb suggested that this new state may be assigned as $\Lambda_b^0(2S)$ resonance, the first radial excitation of the $\Lambda_b^0$ baryon. This resonance was assigned as $\Lambda_b(2S)$ state in the QCD sum rules \cite{Wang2020,Azizi2020}.
Using $^3P_0$ model \cite{Liang2020}, authors considered $\Lambda_b(6072)$ as tentative assignments $\Lambda_b(2S)$, $\Sigma_b(1P)$, and $\rho$-mode excitation of $\Lambda_b(1P)$. Considering the decay widths, they assigned the $\Lambda_b(6072)$ as the lowest $\rho$-mode $\Lambda_b(1P)$ resonance. LHCb Collaboration \cite{LHCb2018} in 2018, observed a new $\Xi_b^-$ resonance with mass, $m_{\Xi_b(6227)^-}=6226.9\pm2.0\pm0.3\pm0.2$ MeV and decay width, $\Gamma_{\Xi_b(6227)^-}=18.1\pm5.4\pm1.8$ MeV in both the $\Lambda_b^0K^-$ and $\Xi_b^0\pi^-$ invariant mass spectra. The resonance is compatible with assignments as $\Xi_b(1P)^-$ \cite{Mao2015,Yang2020} and $\Xi_b(2S)$ \cite{Chen2015}. 

We are using the heavy quark effective theory (HQET) to study the masses of radially excited charm and bottom baryons. The $n=1$ S-wave charm and bottom baryons are used as input to compute the masses of $n=2$ S-wave charm and bottom baryons. The symmetry of HQET parameters is used. The leading order non-perturbative parameters of HQET upto $\frac{1}{m_Q}$ are $\overline{\Lambda}$, $\lambda_1$ and $\lambda_2^Q$. In limit $m_Q\rightarrow\infty$, the $\frac{1}{m_Q}$ term in HQET Lagrangian vanishes. At the order of $m_Q^0$, all hadrons get the contribution to mass from $\overline{\Lambda}$. $\lambda_1$ gives the kinetic energy of the heavy quark and $\lambda_2$ shows the chromomagnetic interaction of the heavy quark. The $\overline{\Lambda}$ parameter comes from the leading term of Lagrangian. We expect it to have a significant contribution to the mass of heavy-light hadrons. The higher order parameters $\lambda_1$ and $\lambda_2$ have a smaller contribution in mass. These parameters are well studied for charm and bottom mesons. Using the data of inclusive semileptonic decay $B\rightarrow Xl\nu_e$ from CLEO \cite{CLeo1996}, the authors in Ref. \cite{Gremm1996} computed $\overline{\Lambda}=0.39\pm0.11$ GeV and $\lambda_1=-0.19\pm0.10$ GeV$^2$. They also computed the bottom and charm quark masses in $\overline{MS}$ scheme as $\overline{m}_b(m_b)=4.45$ GeV and $\overline{m}_c(m_c)=1.28$ GeV. In Ref. \cite{AndreManohar1998,AndreManohar1999}, authors determined  $\lambda_1=-0.27\pm0.10\pm0.04$ GeV$^2$ for $B$ decays. The lattice QCD \cite{KRONFELD2000} was employed to compute the non-perturbative parameters, $\overline{\Lambda}=0.68^{+0.02}_{-0.12}$ GeV and $\lambda_1=-(0.45\pm0.12)$ GeV$^2$. 

There are studies related to these parameters for baryons also. Using sum rules within the framework of HQET in Ref. \cite{DAI1996}, parameters $\overline{\Lambda}_{\Lambda} = 0.79\pm0.05$ GeV for $\Lambda_Q$ baryons and $\overline{\Lambda}_{\Sigma} = 0.96\pm0.05$ for $\Sigma_Q^{(*)}$ baryons were calculated. Also, they computed heavy quark masses to be $m_c=1.43\pm0.05$ GeV and $m_b=4.83\pm0.07$ GeV. The authors in Ref. \cite{Papucci2022} computed the parameters $\overline{\Lambda}_{\Lambda}=0.81$ GeV and $\lambda_{\Lambda,1}=-0.26$ GeV$^2$ using the $m_b=4.71$ \cite{Hoang1999} GeV. Using sum rules \cite{Wang2003}, the parameter $\overline{\Lambda}=0.73\pm0.07$ GeV and $\overline{\Lambda}_{\Sigma}=0.90\pm0.14$ GeV are computed. These non-perturbative parameters can be used to find the masses of excited states. 

This paper is organized in the following order: In sec. 2, a brief overview of the theoretical framework of HQET is given. The HQET Lagrangian is shown with the non-perturbative parameters and mass formulae. In sec. 3, the significance of non-perturbative parameters is analyzed. The role of heavy quark mass is discussed in the hadron using the ratios of mass terms containing parameters $\overline{\Lambda}$ and $\lambda_1$. In sec. 4, the masses of $n=2$ baryons for S-wave are calculated. The strong and semi-electronic decays of charm and bottom baryon decays are studied for $n=1$ and $n=2$ in sec. 5. The strong couplings are also discussed for radially excited states. The conclusions are given in the sec. 6.
\section{Framework}
The hadrons containing a single heavy quark ($c$ or $b$) are studied using HQET. The heavy quark is considered to be much heavier than the light quarks. In the limit $m_Q\rightarrow \infty$, the spins of light quark get decoupled from heavy quark spin. In mesons ($Q\Bar{q}$), the spin of light quark ($s_q$) couples with the orbital angular momentum ($l$) to give a total light spin $s_l=s_q\pm l$. This total light spin ($s_l$) couples with heavy quark spin ($s_Q$) to give the total spin of mesons $J=s_Q\pm s_l$. Spins $s_Q$ and $s_q$ are $\frac{1}{2}$ as quarks are fermions. The total spin $J$ forms a doublet. These doublets are degenerate in the taken limit $m_Q\rightarrow \infty$. 

\begin{figure}[htp]
\centering
\subfloat[Baryons with $s_q=0$ and $J^P=\frac{1}{2}^+$. The flavor $\bar{\textbf{3}}$ representation of SU(3).
\label{3barrep}]{%
\includegraphics[width=4cm]{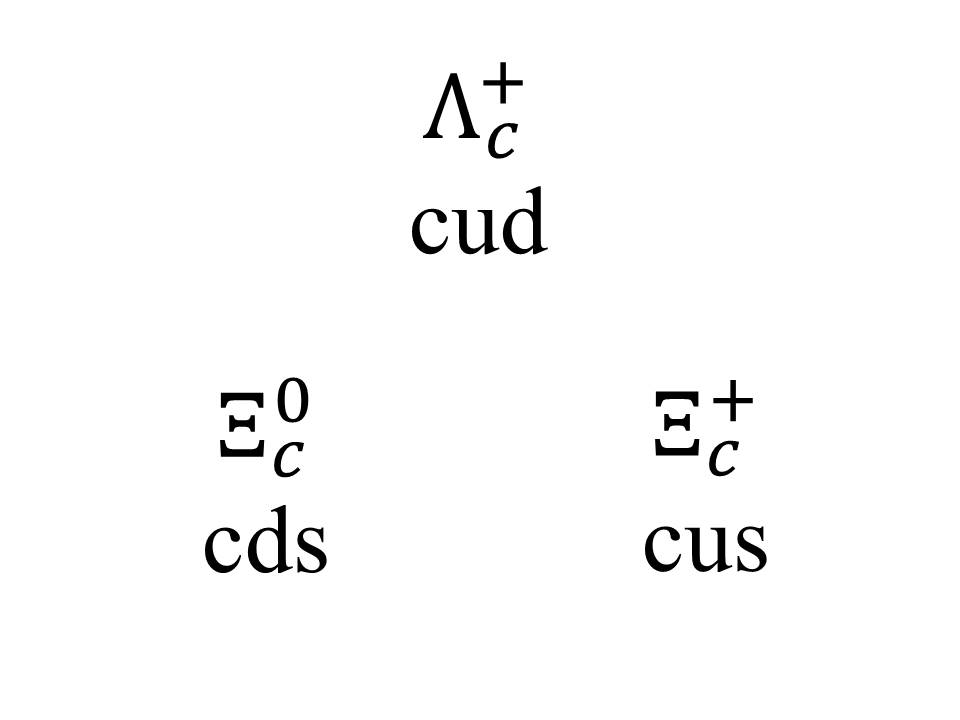}} \hspace{2cm}
\subfloat[Baryons with $s_q=1$ and $J^P=\frac{1}{2}^+$ and $\frac{3}{2}^+$. The baryons with $J^P=\frac{3}{2}^+$ are denoted with $*$ in superscript. The flavor $\textbf{6}$ representation of SU(3). \label{6rep}]{%
    \includegraphics[width=6cm]{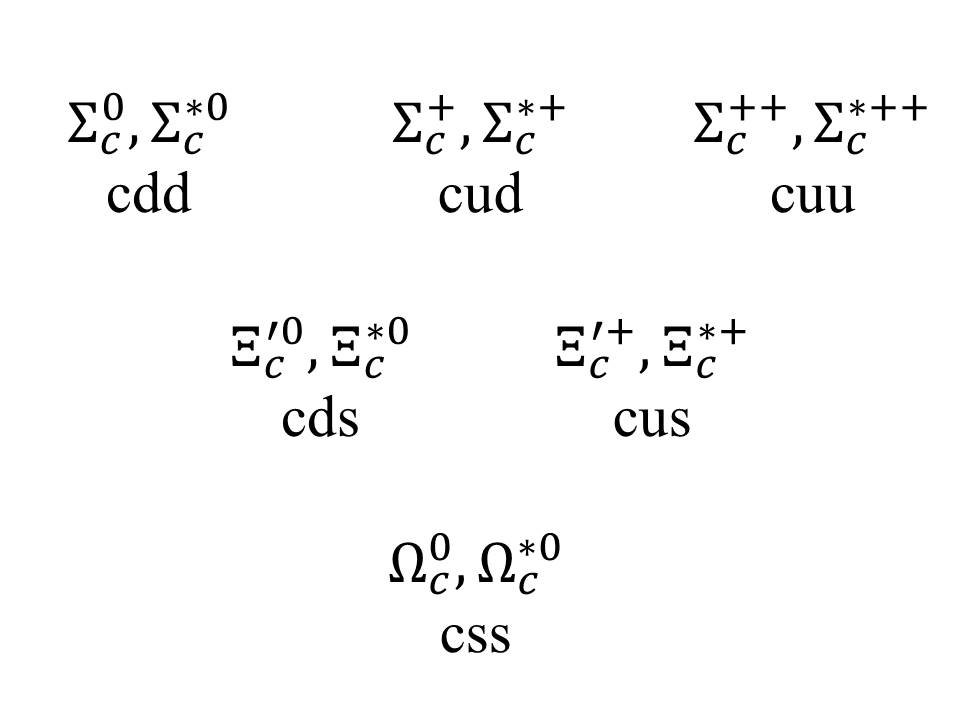}
}
  \caption{The different multiplets of baryons with a heavy charm ($c$) quark. Similar multiplets are for baryons with heavy bottom ($b$) quark. }
  \label{im:surep} 
\end{figure}

For baryons($Qqq$), the spin of light quarks couples to form spin $s_q=0$ and 1. For ground state ($l=0$), spin $s_q$ couples further with heavy quark spin ($s_Q$) to give the total spin $J=\frac{1}{2}$ for $s_q=0$, and $J=\frac{1}{2}$ and $\frac{3}{2}$ for $s_q=1$. As in mesons, the baryons with $s_q=1$ are degenerate in the limit $m_Q\rightarrow \infty$. By taking the effects of heavy quark mass ($m_Q$) to be finite, this degeneracy is broken. The states with $s_q=0$ and $J=\frac{1}{2}$ are denoted by $\Lambda$ ($Qud$) and $\Xi$ ($Qus$ and $Qds$). States with $s_q=1$ and $J=\frac{1}{2}$ are denoted by $\Sigma$ ($Quu$, $Qud$ and $Qdd$), $\Xi^{'}$ ($Qus$ and $Qds$) and $\Omega$ ($Qss$). And states with $s_q=1$ and $J=\frac{3}{2}$ are denoted by $\Sigma^{*}$ ($Quu$, $Qud$ and $Qdd$), $\Xi^{'*}$ ($Qus$ and $Qds$) and $\Omega^{*}$ ($Qss$). The baryons containing the two light quarks can be represented using SU(3) symmetry by $3\otimes3=\overline{3}\oplus6$. These multiplets are shown in Fig. \ref{im:surep}. 
The HQET Lagrangian is written by expanding QCD Lagrangian in terms of heavy quark mass up to $\frac{1}{m_Q}$. The heavy quarks symmetry breaking effects comes from the higher terms of HQET Lagrangian which depends on the heavy quark mass ($m_Q$). The HQET Lagrangian is given as
\begin{align}
\Lagr=\overline{Q}_v(iv.D)Q_v-\overline{Q}_v\frac{D_{\perp}^2}{2m_Q}Q_v-a(\mu)g\overline{Q}_v\frac{\sigma_{\mu\nu}G^{\mu\nu}}{4m_Q}Q_v \label{eq:lag}
\end{align}
where, $D_{\perp}\equiv D^{\mu}-D.v v^\mu$ and $D^{\mu} \equiv \partial^{\mu}-igA^{\mu}$ is the covariant derivative. $v$ is the heavy quark velocity. In the limit $m_Q\rightarrow\infty$, the heavy quark velocity is the velocity of hadron. $G^{\mu\nu}$ is the gluon field strength tensor. Only the first term survives in the limit $m_Q\rightarrow\infty$. Higher terms contain $\frac{1}{m_Q}$ factor and thus break the heavy quark symmetry. $Q_v$ is the effective heavy field. The field $Q_v$ can be written in the form of $H_v^{(Q)}$, which is the covariant representation of heavy field.
\begin{align}
    \text{For ground state mesons}:~~
    H_v^{(Q)}=\frac{1+\slashed{v}}{2}[\slashed{P}_v^{*(Q)}+i P_v^{(Q)}\gamma_5]
\end{align}
where, $\frac{1+\slashed{v}}{2}$ is the projection operator of the meson, $P_v^{*(Q)}$ is the vector field and $P_v^{(Q)}$ is the pseudoscalar field that annihilates the meson multiplet with $s_l=1/2$. Similar fields can be written for baryons also. The $\Bar{3}$ multiplet can be represented as an antisymmetric matrix $B_{\bar{3}}$ \cite{TMYan1992} or a vector T \cite{pcho1992} with $s_l=0$. The $6$ multiplet is represented by the $S$ field \cite{pcho1992}, describing both $J=\frac{1}{2}$ and $\frac{3}{2}$ baryons for $s_l=1$. 
\begin{align}
    T_i=\frac{1+\slashed{v}}{2}\begin{pmatrix}
        \Xi_c^0 & -\Xi_c^+ & \Lambda_c^+
    \end{pmatrix}_i=\frac{1}{2}\epsilon_{ijk}(B_{\bar{3}})_{jk} \label{eq:tfield}\\
    S^{ij}_{\mu}=\frac{1}{\sqrt{3}}(\gamma_{\mu}+v_{\mu})\gamma_5\frac{1+\slashed{v}}{2}B^{ij}_6+\frac{1+\slashed{v}}{2}B^{*ij}_{6\mu}
    \label{eq:sfield}
\end{align}
where, matrices $B_{\bar{3}}$ and $B_6$ are given below as defined in \cite{TMYan1992}. The matrix $B^*_{6\mu}$ is similar to $B_6$. $B^*_{6\mu}$ is a Rarita-Schwinger vector-spinor field for spin $J=\frac{3}{2}$ baryon.
\begin{align}
    B_{\bar{3}} & = \begin{bmatrix}
        0 & \Lambda_c^+ & \Xi_c^+\\
        -\Lambda_c^+ & 0 & \Xi_c^0\\
        - \Xi_Q^+ & - \Xi_Q^0 & 0
    \end{bmatrix} \label{eq:b3matrix}\\
    B_6 & = \begin{bmatrix}
        \Sigma_c^{++} & \frac{1}{\sqrt{2}}\Sigma_c^+ & \frac{1}{\sqrt{2}}\Xi_c^{+'}\\
         \frac{1}{\sqrt{2}}\Sigma_c^+ & \Sigma_c^{0} & \frac{1}{\sqrt{2}}\Xi_c^{0'}\\
         \frac{1}{\sqrt{2}}\Xi_c^{+'} & \frac{1}{\sqrt{2}}\Xi_c^{0'} & \Omega_c^0
    \end{bmatrix} \label{eq:b6matrix}
\end{align}
The masses of the hadrons can be obtained by using heavy quark symmetry. All hadrons containing a single heavy heavy quark (Q) are degenerate at order $m_Q$, and have mass $m_Q$. At order $m_Q^0$, the hadron masses get a contribution from the first term of the Lagrangian as shown in Eq. \eqref{eq:lag}. 
\begin{align}
     \overline{\Lambda}&\equiv\frac{1}{2}\matrixel{ H^{(Q)}}  {H_0}  {H^{(Q)}}
     \label{eq:Lambdabar}\\
    2\lambda_1&=-\matrixel{H^{(Q)}}{\overline{Q}_v D_{\perp}^2 Q_v}{H^{(Q)}} \label{eq:lamb1}\\
    16(S_Q.s_l)\lambda_2^Q&= a(\mu) \matrixel{H^{(Q)}}{\overline{Q}_v g\sigma_{\alpha\beta} G^{\alpha\beta} Q_v}{H^{(Q)}}
    \label{eq:lamb2}
\end{align}    
where in Eq. \eqref{eq:Lambdabar}, $H_0$ is the $\frac{1}{m_Q^0}$ order term of the Hamiltonian of HQET obtained from the first term of Lagrangian. $H^{(Q)}$ is the hadron states in the effective theory with $v=(1,\bf{0})$. $\overline{\Lambda}$ is a HQET parameter. It has the same value for all particles in a spin-flavor multiplet. We denote $\overline{\Lambda}_H$ for $D(D^*)$ and $B(B^*)$ mesons, $\overline{\Lambda}_{\Lambda}$ for $\Lambda_{c(b)}$ baryons, $\overline{\Lambda}_{\Xi}$ for $\Xi_{c(b)}$ baryons, $\overline{\Lambda}_{\Sigma}$ for $\Sigma_{c(b)}^{(*)}$ baryons, $\overline{\Lambda}_{\Xi'}$ for $\Xi^{'(*)}_c(b)$ baryons and, $\overline{\Lambda}_{\Omega}$ for $\Omega_{c(b)}^{(*)}$ baroyns. In Eq. \eqref{eq:lamb1}, $\lambda_1$ parameter is independent of $m_Q$ but is different for different multiplets of baryons. In Eq. \eqref{eq:lamb2}, $\lambda_2^Q$ parameter depends on $m_Q$ through the dependence of $a(\mu)$ on $m_Q$. In the leading logarithmic approximation
\begin{align}
    a(\mu)=\left[\frac{\alpha_s(m_Q)}{\alpha_s(\mu)} \right]^{9/(33-2N_q)}\label{eq:alpha}
\end{align}
where, $N_q$ is the number of light quark flavors. The $\lambda_2$ matrix element transform like $s_Q.s_l$ under the spin symmetry, as $\overline{Q}_v g\sigma_{\alpha\beta} G^{\alpha\beta} Q_v$ has the same transformation property. The operator $\overline{Q}_v\sigma Q_v$ is the heavy quark spin \cite{manohar_wise_2000}. Using $s_Q.s_l=(J^2-s_Q^2-s_l^2)/2$, the mass equations for hadrons can be written. 
The mass formula for heavy hadron H containing heavy quark Q in terms of non-perturbative HQET parameters is given below
\begin{align}
    m_{H^{(Q)}}=m_Q+\overline{\Lambda}-\frac{\lambda_{H,1}}{2m_Q}\pm n_{\mp}\frac{\lambda_{H,2}}{2m_Q}
    \label{eq:massH}
\end{align}
where, $n_{\pm}=2J_{\pm}+1$. The $H$ in subscript represents the dependence of parameters $\lambda_1$ and $\lambda_2$ on the multiplet of hadrons. For ground state charm mesons $H=(D,D^*)$ and bottom meson $H=(B,B^*)$. The $\bar{3}$ baryons ($\Lambda_Q$ and $\Xi_Q$) are singlet states with $J=\frac{1}{2}$. So, $n=0$ for $\Lambda_Q$ and $\Xi_Q$ baryons. The $6$ baryons ($\Sigma^{(*)}_Q, \Xi^{'(*)}_Q$ and $\Omega^{(*)}_Q$) with $J=\frac{1}{2}$ form a doublet with $J=\frac{3}{2}$ baryons. 
The SU(3) symmetry is broken for $u$, $d$ and $s$ quarks as, $s$ quark is much heavier than $u$ and $d$ quarks. So, different parameters are used for baryons containing a different number of strange quarks. We have used $\Lambda (\Sigma)$, $\Xi (\Xi')$, and $\Omega$ in subscript to denote the presence of zero, one, and two strange quarks in hadrons, respectively. While the $\bar{\Lambda}$ and $\lambda_1$ parameters can be computed from the above mass equations. $\lambda_2$ parameter is different as it depends on heavy flavor of quark through Eq. \eqref{eq:alpha}, and we will calculate it with the below equation
 \begin{align}
     \lambda_2^Q=\frac{1}{8}\left(m_{\Sigma_Q^*}^2-m_{\Sigma_Q}^2\right)\label{eq:lamb2mass}
 \end{align}
 This equation is coming from the last term of Lagrangian Eq. \eqref{eq:lag}.  Also, the difference of $\lambda_2^b$ and $\lambda_2^c$ is coming from the relation in Eq. \eqref{eq:alpha} as given in Ref. \cite{AMOROS1997}
\begin{align}
    \lambda_2^b=\lambda_2^c\left(\frac{\alpha(m_b)}{\alpha(m_c)}\right)^{9/25}\label{eq:lam2corr}
\end{align}
\section{Analysis of non-perturbative parameters}
The non-perturbative parameters ($\overline{\Lambda}$, $\lambda_1$, $\lambda_2$) of HQET are useful to find masses, decay width, etc \cite{manohar_wise_2000}. These parameters are shown in the above mass Eqs. \eqref{eq:massH}. The masses of heavy-light hadrons depend on the nature of these non-perturbative parameters. These parameters are well-studied for heavy-light mesons, as discussed in Sec 1. Here, we have computed the values of all $\overline{\Lambda}$, $\lambda_1$, and $\lambda_2^{Q}$ parameters using masses of charm and bottom mesons taken from PDG \cite{PDG2021} and from Ref. \cite{Pallavi2018} as given in Table \ref{tab:Smesons}. The $\overline{\Lambda}$ and $\lambda_1$ are calculated with average masses of spin partners of charm and bottom mesons according to Eq. \eqref{eq:massH}. The $\lambda_2^Q$ has been computed using Eq. \eqref{eq:lamb2mass}.
\begin{table}[t]
    \centering
    \begin{tabular*}{\textwidth}{@{\extracolsep{\fill}}ccccccc}\toprule
    n & $J^P$ & $c\Bar{q}$  & $c\Bar{s}$ & $b\Bar{q}$ & $b\Bar{s}$\\ \midrule 
    \multirow{2}{*}{1} & $0^-$ & $1869.66\pm0.05$ & $1968.35\pm0.07$ & $5279.34\pm0.12$ & $5366.92\pm0.10$ \\  
                       & $1^-$ & $2010.26\pm0.05$ & $2112.2\pm0.4$ & $5324.71\pm0.21$ & $5415.8\pm1.5$\\\midrule
    \multirow{2}{*}{2} & $0^-$ & $2549\pm19$ & $2591\pm6\pm7$ & $5932\pm13$ \cite{Pallavi2018} & $6029\pm5$ \cite{Pallavi2018} \\ 
                       & $1^-$ & $2627\pm10$ & $2714\pm5$ & $5957\pm13$ \cite{Pallavi2018} & $6056\pm5$ \cite{Pallavi2018} \\  \bottomrule
     \end{tabular*}
    \caption{The S-wave charm and bottom  mesons are shown for both $n=1$ and 2. The masses are taken from PDG \cite{PDG2021} and Ref. \cite{Pallavi2018}. All masses are in MeV.}
    \label{tab:Smesons}
\end{table}
These parameters are calculated using $m_c=1270$ MeV and $m_b=4180$ MeV as shown in Table \ref{tab:Mesonpara}, and with $m_c=1290$ MeV and $m_b=4670$ MeV as shown in Table \ref{tab:Mpara2}. We find a drastic difference between the values of HQET parameters up to $\frac{1}{m_{Q}}$ corrections by changing the heavy quark masses ($m_Q$) shown in Tables \ref{tab:Mesonpara} and \ref{tab:Mpara2}. This demands a further investigation into the nature of these parameters and their contribution to the  masses. The values of parameters for mesons estimated by previous studies are reproduced by taking the heavy quark masses $m_c=1290$ MeV and $m_b=4670$ MeV \cite{Alka2014}. We can look at the contributions of mass terms in Eqs. \eqref{eq:massH} by changing the mass of heavy quarks.
\begin{table}[t]
    \centering
    \begin{tabular*}{\textwidth}{@{\extracolsep{\fill}}cccc}\toprule
    \multicolumn{4}{c}{$n=1$}\\\midrule
    $\overline{\Lambda}_{H}$ & $\lambda_{1}$ & $\lambda^c_{2}$ & $\lambda^b_{2}$ \\ 
    $1320$ & $1.56\times10^6$ & $6.82\times10^4$ & $6.0\times10^4$ \\ \midrule
    $\overline{\Lambda}_{H,s}$ & $\lambda_{1,s}$ & $\lambda^c_{2,s}$ & $\lambda^b_{2,s}$\\
    $1405$ & $1.52\times10^6$ & $7.34\times10^4$ & $6.59\times10^4$\\\midrule
    \multicolumn{4}{c}{$n=2$}\\\midrule
    $\overline{\Lambda}_{H}$ & $\lambda_{1}$ & $\lambda^c_{2}$ & $\lambda^b_{2}$ \\ 
    $1959$ & $1.58\times10^6$ & $5.05\times10^4$ & $3.72\times10^4$ \\ \midrule
    $\overline{\Lambda}_{H,s}$ & $\lambda_{1,s}$ & $\lambda^c_{2,s}$ & $\lambda^b_{2,s}$\\
    $2068$ & $1.66\times10^6$ & $8.16\times10^4$ & $4.08\times10^4$\\\bottomrule
    \end{tabular*}
    \caption{Computed values of non-perturbative parameters for mesons in HQET using meson masses given in Table \ref{tab:Smesons} and heavy quark masses ($m_c=1270$ MeV and $m_b=4180$ MeV) given in PDG \cite{PDG2021}. The $\overline{\Lambda}$ parameters are in MeV, $\lambda_1$ and $\lambda_2$ parameters are in MeV$^2$. }
    \label{tab:Mesonpara}
\end{table}
\begin{table}[t]
    \centering
    \begin{tabular*}{\textwidth}{@{\extracolsep{\fill}}cccc}\toprule
    \multicolumn{4}{c}{$n=1$}\\\midrule
    $\overline{\Lambda}_{H}$ & $\lambda_{1}$ & $\lambda^c_{2}$ & $\lambda^b_{2}$ \\
    $627$ & $-0.15\times10^6$ & $6.82\times10^4$ & $6.0\times10^4$ \\ \midrule
    $\overline{\Lambda}_{H,s}$ & $\lambda_{1,s}$ & $\lambda^c_{2,s}$ & $\lambda^b_{2,s}$\\
     $713$ & $-0.19\times10^6$ & $7.34\times10^4$ & $6.59\times10^4$\\\midrule
    \multicolumn{4}{c}{$n=2$}\\\midrule
    $\overline{\Lambda}_{H}$ & $\lambda_{1}$ & $\lambda^c_{2}$ & $\lambda^b_{2}$ \\
    $1266$ & $-0.13\times10^6$ & $5.05\times10^4$ & $3.72\times10^4$ \\ \midrule
    $\overline{\Lambda}_{H,s}$ & $\lambda_{1,s}$ & $ \lambda^c_{2,s}$ & $\lambda^b_{2,s}$\\
    $1373$ & $-0.05\times10^6$ & $8.16\times10^4$ & $4.08\times10^4$\\\bottomrule
    \end{tabular*}
    \caption{Computed values of non-perturbative parameters for mesons in HQET using masses given in Table \ref{tab:Smesons} and heavy quark masses ($m_c=1290$ MeV and $m_b=4670$ MeV) from Ref. \cite{Alka2014}.}
    \label{tab:Mpara2}
\end{table}

For $m_c=1270$ MeV and $m_b=4180$ MeV, the term containing $\lambda_1$ in $D$ and $D^*$ mesons has a contribution of 615.35 MeV, and the term containing $\lambda_2^c$ in $D$ meson has a contribution of -105.46 MeV and in $D^*$ meson of 35.153 MeV. For terms containing $\lambda_1$ in $B$ and $B^*$ mesons, the contribution is about 186.96 MeV. In terms containing $\lambda_2$ in $B$ and $B^*$ meson, the contribution is 34.00 MeV and 11.33 MeV, respectively. We find a significant reduction in mass term contributions of bottom mesons. This may be due to the difference in heavy quark masses ($m_c$ and $m_b$). Thus, higher order contributions decrease with an increase in the mass of heavy quarks. 

Analyzing the contributions to hadron mass by changing the quark masses to $m_c=1290$ MeV and $m_b=4670$ MeV will give us the behavior of terms containing the non-perturbative  parameters. The $\overline{\Lambda}$ parameter in $D$ meson reduced from 1320 MeV to 627 MeV. The same behavior is followed for other states and radially excited states also. The contribution of terms containing $\lambda_1$ in $D$ and $D^*$ is 57.36 MeV and, in $B$ and $B^*$ is 15.84 MeV. The analysis is simple for terms containing $\lambda_2$ parameter. The $\lambda_2$ gives the hyperfine splitting, i.e., mass difference between $J^P=0^-$ and $J^P=1^-$ states of mesons. The contribution of mass terms containing $\lambda_2$ remains the same with change of heavy quark mass.

\begin{figure}[htp]
    \centering
\subfloat[For $n=1$ $D$ and $B$ mesons]{%
    \includegraphics[width=\linewidth]{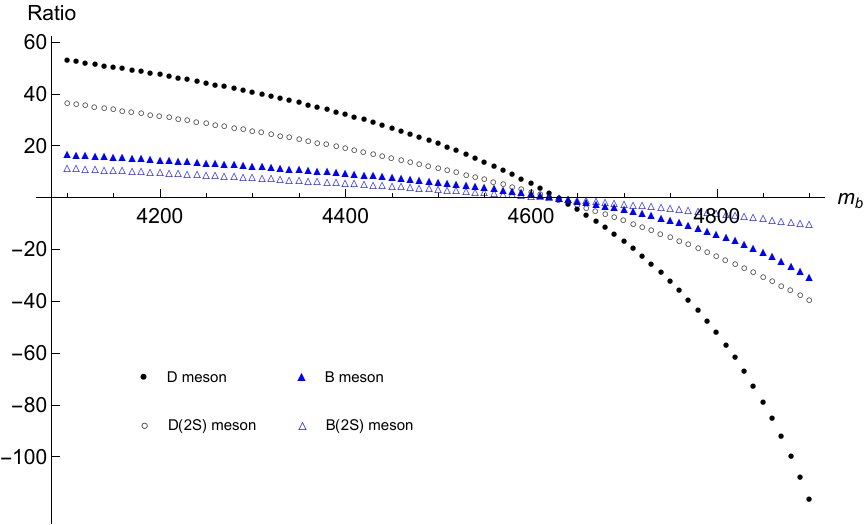}%
    }\\
\subfloat[For $n=1$ $\Lambda_Q$ baryons]{%
    \includegraphics[width=\linewidth]{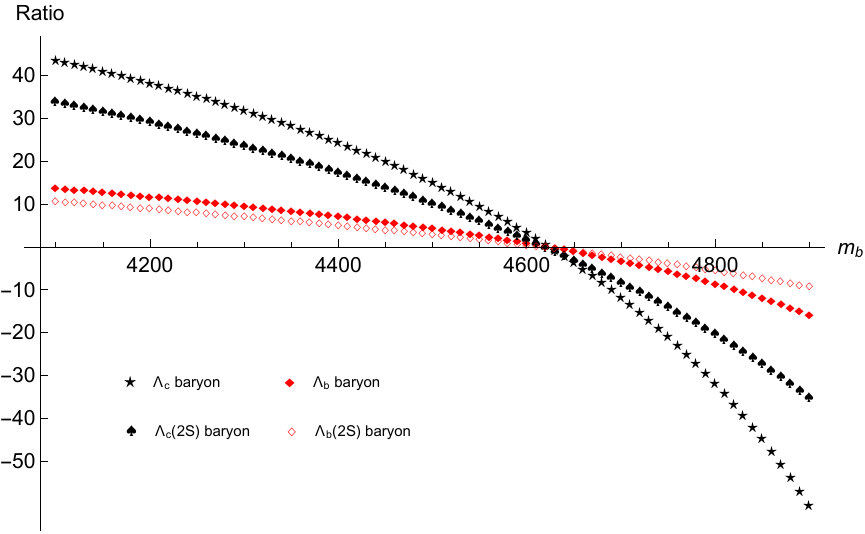}%
    }
    \caption{Ratio (in \%age) of mass terms $\frac{\lambda_1}{2m_Q}$ and $\overline{\Lambda}$ with $m_b$ (MeV).}
    \label{fig:ratiomb}
\end{figure}

Further, the ratio of the mass terms containing parameters $\lambda_1$ and $\overline{\Lambda}$ are analyzed. As $\lambda_1$ represents the higher order correction, we expect the correction coming from it to be small. When the heavy quark masses are taken as $m_c=1270$ MeV and $m_b=4180$ MeV, for $D$ and $B$ mesons the ratio $\frac{\lambda_1}{2m_c}\frac{1}{\overline{\Lambda}}=46.6\%$ and $\frac{\lambda_1}{2m_b}\frac{1}{\overline{\Lambda}}=14.2\%$ respectively. For heavy quark masses $m_c=1290$ MeV and $m_b=4670$ MeV, the ratio changes to $\frac{\lambda_1}{2m_c}\frac{1}{\overline{\Lambda}}=9.1\%$ and $\frac{\lambda_1}{2m_b}\frac{1}{\overline{\Lambda}}=2.52\%$. The values of the ratio given above are absolute values. These ratios show that the masses of heavy quarks given in PDG \cite{PDG2021} may not be used for these calculations. We have plotted in Fig. \ref{fig:ratiomb}, the above ratios with the change of bottom quark mass ($m_b$) and keeping the mass of charm quark to be $m_c=$ 1290 MeV. To plot these ratios, the formulae of spin average masses of both charm and bottom hadrons are solved simultaneously for parameters $\overline{\Lambda}$ and $\lambda_1$. As these parameters are the same for both charm and bottom hadrons, they can be solved simultaneously. The bottom quark mass $m_b$ is varied with an increment of 10 MeV in the range of 4100 MeV to 4900 MeV, while solving the simultaneous equations. The sets of parameters ($\overline{\Lambda}$ and $\lambda_1$) from solving these equations are used to find the ratios $\frac{\lambda_1}{2m_Q}$$\frac{1}{\overline{\Lambda}}$, which are plotted with $m_b$ as shown in Fig. \ref{fig:ratiomb}.   This simply shows that the parameter $\lambda_1$ changes sign at $m_b=4620$ MeV. Most of the theoretical predictions and their averages give $\lambda_1$ to be negative for hadrons. Thus, the mass of the bottom quark $m_b$ inside the hadrons may be greater than 4620 MeV. 
The ratio $\frac{\lambda_1}{2m_c}\frac{1}{\overline{\Lambda}}$ goes below the $m_b$ axis at the same point in both mesosns and baryons shown in Fig. \ref{fig:ratiomb}. This may indicate that the heavy quark behaves almost the same in both mesons and baryons. The mass formulae in Eq. \eqref{eq:massH} assumes the nature of heavy quark same in all hadrons. Thus, Fig. \ref{fig:ratiomb} confirms this assumption. 
\section{Masses of 2S-wave baryons}
\begin{figure}
\centering
    \subfloat[\label{fig:S3_c_l1}]{%
    \includegraphics[width=0.4\linewidth]{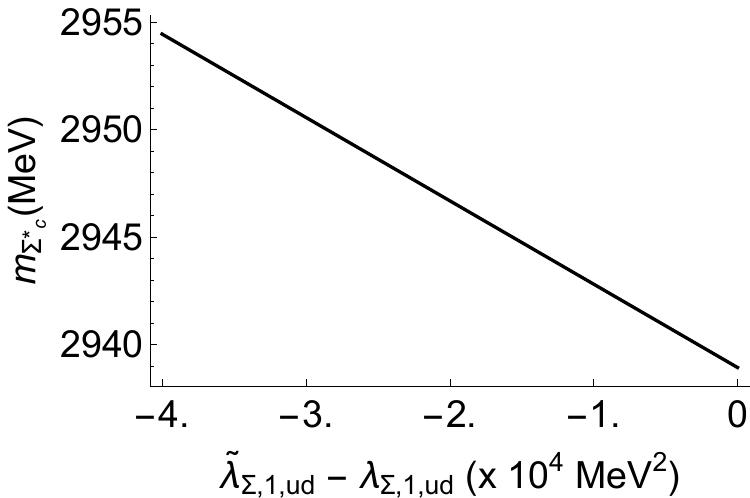}
    }\hspace{1cm}
    \subfloat[\label{fig:S3_c_l2}]{%
    \includegraphics[width=0.4\linewidth]{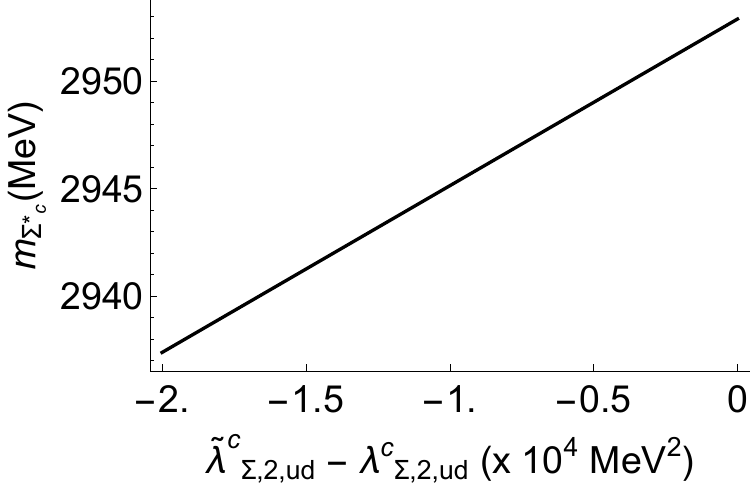}
    }\\
    \subfloat[\label{fig:S3_b_l1}]{%
    \includegraphics[width=0.4\linewidth]{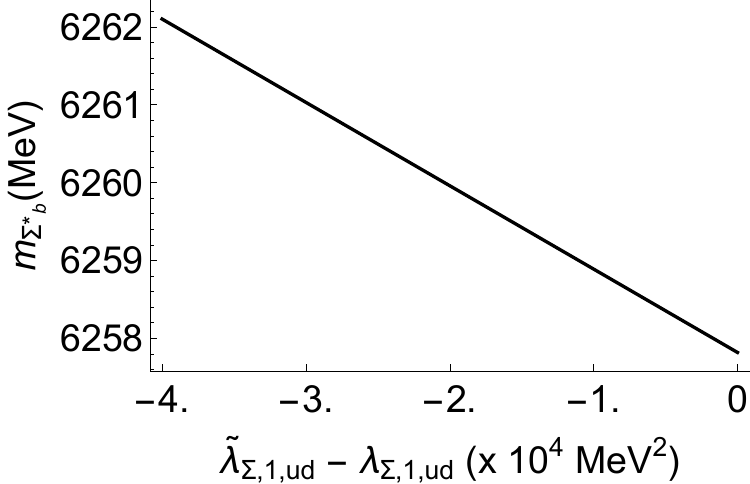}
    }\hspace{1cm}
    \subfloat[\label{fig:S3_b_l2}]{%
    \includegraphics[width=0.4\linewidth]{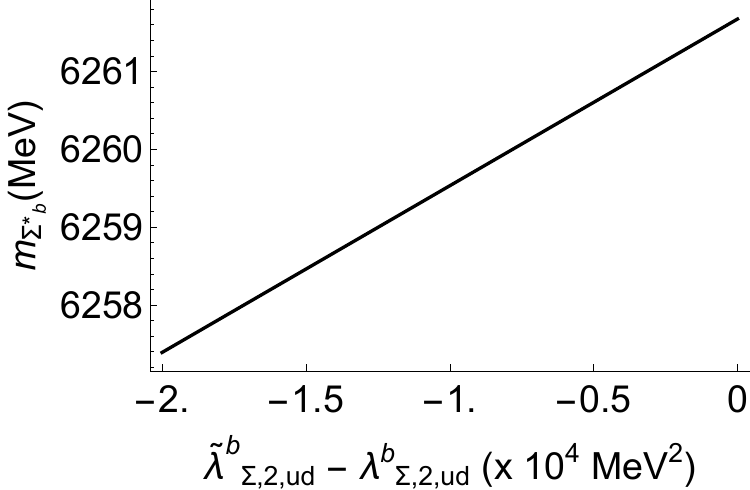}
    }\\
    \caption{The variation of masses of $\Sigma^*_Q(2S)$ with both $
\widetilde{\lambda}_{\Sigma,1,ud}-\lambda_{\Sigma,1,ud}$ and $\widetilde{\lambda}_{\Sigma,2,ud}-\lambda_{\Sigma,2,ud}$}
\label{fig:S3_l1_l2}
\end{figure}

The masses of 1S heavy baryons shown in Table \ref{tab:Swave} are taken from PDG \cite{PDG2021} and Ref. \cite{Omegab32} given in Table \ref{tab:paran1}. The non-perturbative parameters are calculated from the mass formulae given in Eqn. \eqref{eq:massH}. The heavy quark masses used are $m_c=1290$ MeV and $m_b=4670$ MeV. The same heavy quark masses are taken for mesons. To calculate masses for $n=2$ S-wave charm and bottom baryons, the following HQET symmetry is used: 
\begin{align}
    \widetilde{\overline{\Lambda}}_{\Xi(\Xi')} - \widetilde{\overline{\Lambda}}_{\Lambda(\Sigma)} &\approx \overline{\Lambda}_{\Xi(\Xi')} - \overline{\Lambda}_{\Lambda(\Sigma)}\label{eq:symm1}
\end{align}
\begin{table}[t]
    \centering
    \begin{tabular*}{\textwidth}{@{\extracolsep{\fill}}cccc}\toprule
    $J^P$ & Baryons  & $Q=c$ & $Q=b$  \\\midrule
    \multirow{5}{*}{$\frac{1}{2}^+$} & $\Lambda$ & $2286.46\pm0.14$ & $5619.60\pm0.17$\\
                  & $\Xi$ & $2469.08\pm0.18$ & $5794.45\pm0.40$ \\
                  & $\Sigma$ & $2453.54\pm0.15$ & $5813.10\pm0.18$\\
                  & $\Xi^{'}$ & $2578.45\pm0.35$ & $5935.02\pm0.05$\\
                  & $\Omega$ & $2695.2\pm1.7$ & $6046.1\pm1.7$ \\\midrule
    \multirow{3}{*}{$\frac{3}{2}^+$} & $\Sigma^*$ & $2518.13\pm0.8$ & $5832.53\pm0.20$\\
                  & $\Xi^{'*}$ & $2645.63\pm0.20$ & $5953.82\pm0.31$\\
                  & $\Omega^*$ & $2766.0\pm2.0$ & $6082\pm20$*\\ \bottomrule
     \end{tabular*}
    \caption{The masses of S-wave charm and bottom  baryons are shown in MeV. The mass of $\Omega_b^*$ is taken from Ref. \cite{Omegab32} and all other masses are taken from PDG.}
    \label{tab:Swave}
\end{table}
The Eq. \eqref{eq:symm1} gives the difference of parameters for strange and non-strange hadrons is similar for higher radially excited states. As parameter $\widetilde{\lambda}_{\Lambda,1}$ and $\widetilde{\lambda}_{\Lambda,2}$ comes from higher order correction term $\left(\frac{1}{m_Q}\right)$. To estimate the value of these parameters for higher excited states, the same case in mesons can be used. As mesons are well studied sector under the same framework, we can get a better understanding of these non-perturbative parameters for $n=2$ case. Table \ref{tab:Mesonpara} suggests that on going from $n=1$ to 2 the parameter $\overline{\Lambda}$ is increased as expected, whereas the parameters $\lambda_1$ and $\lambda_2$ are decreased. 
\begin{table}[t]
    \centering
    \begin{tabular*}{\textwidth}{@{\extracolsep{\fill}}cccc}\toprule
    \multicolumn{4}{c}{$n=1$}\\\midrule
    $\overline{\Lambda}_{\Lambda}$ & $\lambda_{\Lambda,1}$ & $\overline{\Lambda}_{\Xi}$ & $\lambda_{\Xi,1}$\\
    $931$ & $-0.17\times10^6$ & $1103$ & $-0.19\times10^6$\\\midrule
    $\overline{\Lambda}_{\Sigma}$ & $\lambda_{\Sigma,1}$ & $\lambda^{c}_{\Sigma,2}$ & $\lambda^{b}_{\Sigma,2}$\\
    $1136$ & $-0.18\times10^6$ & $4.01\times10^4$ & $2.83\times10^4$\\\midrule
    $\overline{\Lambda}_{\Xi'}$ & $\lambda_{\Xi',1}$ & $\lambda^{c}_{\Xi',2}$ & $\lambda^{b}_{\Xi',2}$\\
    $1256$ & $-0.20\times10^6$ & $4.39\times10^4$ & $2.79\times10^4$\\\midrule
    $\overline{\Lambda}_{\Omega}$ & $\lambda_{\Omega,1}$ & $\lambda^{c}_{\Omega,2}$  & $\lambda^{b}_{\Omega,2}$\\
    $1380$ & $-0.19$ x $10^6$ & $4.83\times10^4$ & $5.44\times10^4$\\\bottomrule
    \end{tabular*}
    \caption{Computed parameters for $n=1$ S-wave baryons. The parameters $\overline{\Lambda}$ are in units MeV. $\lambda_1$ and $\lambda_2$ are in MeV$^2$ units.}
    \label{tab:paran1}
\end{table}
Extending this behavior to baryons using the Eq. \eqref{eq:ldcud}, which is the difference of mass of $\Lambda_c$ (given by Eq. \eqref{eq:massH}) for $n=2$ and 1. 
\begin{align}
    m_{\Lambda_c(2S)}-m_{\Lambda_c}=\widetilde{\overline{\Lambda}}_{\Lambda}-\overline{\Lambda}_{\Lambda}-\frac{\widetilde{\lambda}_{\Lambda,1}-\lambda_{\Lambda,1}}{2m_c} 
    \label{eq:ldcud}
\end{align}
where, $m_{\Lambda_c(2S)} = 2766.6\pm2.4$ MeV is the mass of radially excited $\Lambda_c$ with $n=2$ taken from PDG \cite{PDG2021}. The parameter $\lambda_{1}$ in the case of mesons does not change significantly from $n=1$ and 2.  We take $\widetilde{\lambda}_{\Lambda,1}=-0.19*10^6$ which is smaller than $\lambda_{\Lambda,1}$ from Table \ref{tab:paran1}. The parameter $\widetilde{\overline{\Lambda}}_{\Lambda}$ can be calculated for $n=2$ from Eq. \eqref{eq:ldcud} as shown in Table \ref{tab:paran2}. Now using these parameters in the Eq. \eqref{eq:ldbud}, we get mass for $m_{\Lambda_b(2S)}$, shown in Table \ref{tab:Swaven2}. 
\begin{figure}
    \centering
    \subfloat[Mass of $\Lambda_c(2S)$ with difference of $\lambda_1$ for $n=2$ and 1 \label{fig:L_c_l1}]{%
    \includegraphics[width=0.41\linewidth]{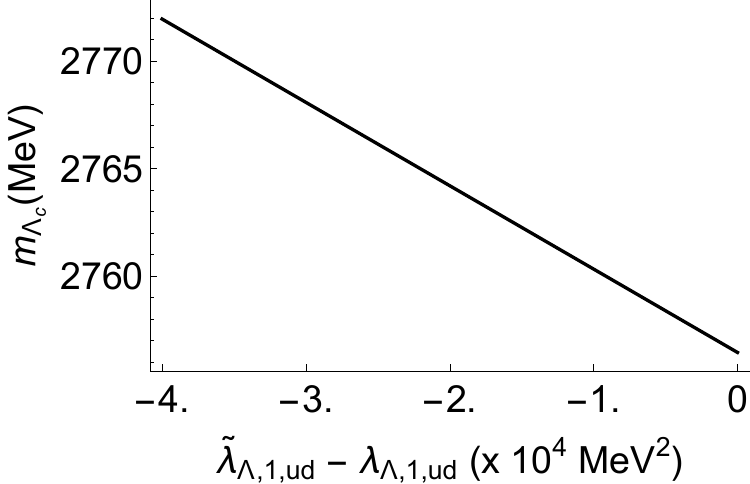}
    }\hspace{1cm}
    \subfloat[Mass of $\Lambda_b(2S)$ with difference of $\lambda_1$ for $n=2$ and 1 \label{fig:L_b_l1}]{%
    \includegraphics[width=0.41\linewidth]{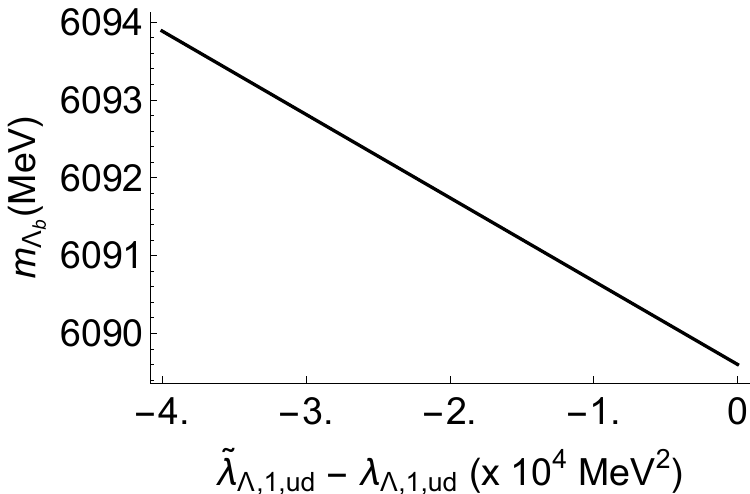}
    }\hspace{1cm}
    \caption{The variation of masses of $\Lambda_Q(2S)$ with $\lambda_{\Lambda,1,ud}$}
    \label{fig:L_l1}
\end{figure}
In Fig. \ref{fig:L_l1}, the mass of $\Lambda_Q(2S)$ is varying with the difference of $\widetilde{\lambda}_{\Lambda,1} - \lambda_{\Lambda,1}$ parameters. The variation of the difference in parameters shows the dependence and effect of these parameters on the masses of radially excited baryons. The mass of $\overline{\Lambda}_Q$ baryon is affected slightly by the variation of parameters from $n=1$ to 2. The above behavior of excited baryon masses is expected in heavy quark symmetry. The same behavior is also expected for the parameter $\lambda_{2}$ as shown in Fig. \ref{fig:S3_l1_l2} for the $\Sigma^*_Q$ baryon. The $\Sigma^*_Q$ has $J^P=\frac{3}{2}^+$, this make a positive contribution of $\lambda_2$ parameter as given in Eq. \eqref{eq:massH} and \eqref{eq:massH}, thus a positive slope in Figs. \ref{fig:S3_c_l2} and \ref{fig:S3_b_l2}. The Figs. \ref{fig:S3_c_l1} and \ref{fig:S3_b_l1} are similar to Figs. \ref{fig:L_c_l1} and \ref{fig:L_b_l1} as the masses of baryons are changing within the same range. The masses of other baryons $\Xi_Q$ and $\Omega_Q$ also show similar results. The dependence on parameters of baryon masses justifies the heavy quark symmetry. The values of parameters $\lambda_1$ and $\lambda_2$ are taken by keeping these considerations. 
\begin{align}
    m_{\Lambda_b(2S)}-m_{\Lambda_b}=\widetilde{\overline{\Lambda}}_{\Lambda}-\overline{\Lambda}_{\Lambda}-\frac{\widetilde{\lambda}_{\Lambda,1}-\lambda_{\Lambda,1}}{2m_b} \label{eq:ldbud}
\end{align}
For $\Xi_c$ and $\Xi_b$, we invoke the symmetry given by Eq. \eqref{eq:symm1} to get $\widetilde{\overline{\Lambda}}_{\Xi}=1574$ MeV as shown in Table \ref{tab:paran2}. Also, we take $\widetilde{\lambda}_{\Xi,1}=-0.21*10^6$ MeV$^2$, which preserves the difference of $\widetilde{\lambda}_{\Lambda,1}$ and $\lambda_{\Lambda,1}$. Using Eqs. \eqref{eq:ldcbs} equations similar to Eqs. \eqref{eq:ldcud} and \eqref{eq:ldbud}, we can now get masses for $\Xi_c(2S)$ and $\Xi_b(2S)$ for $J^P=\frac{1}{2}^+$ as shown in Table \ref{tab:Swaven2}. 
\begin{align}
     m_{\Xi_Q(2S)}-m_{\Xi_Q}=\widetilde{\overline{\Lambda}}_{\Xi}-\overline{\Lambda}_{\Xi}-\frac{\widetilde{\lambda}_{\Xi,1}-\lambda_{\Xi,1}}{2m_Q} \label{eq:ldcs}\\
     \label{eq:ldcbs}
\end{align}
For the \textbf{6} multiplet of SU(3) representation we take mass $m_{\Sigma_c(2S)}=2901$ MeV from Ref. \cite{ebert2011} as input. We also take $\widetilde{\lambda}_{\Sigma,1}=-0.20*10^6$ MeV$^2$ and $\widetilde{\lambda}^c_{\Sigma,2}=4.01*10^4$ MeV$^2$. We have used the same value of $\widetilde{\lambda}^c_{\Sigma,2}$ as their $n=1$ counterparts from Table \ref{tab:paran1}. Then we can use the Eq. \eqref{eq:sdcud1} to find $\widetilde{\overline{\Lambda}}_{\Sigma}$, shown in Table \ref{tab:paran2}. 
\begin{align}
    \overline{m}_{\Sigma_c(2S)}-\overline{m}_{\Sigma_c}=\widetilde{\overline{\Lambda}}_{\Sigma}-\overline{\Lambda}_{\Sigma}-\frac{\widetilde{\lambda}_{\Sigma,1}-\lambda_{\Sigma,1}}{2m_c} \label{eq:sdcud1}
\end{align}
where $\bar{m}_{\Sigma_c}$ is the average mass of spin partners $\Sigma_c$ and $\Sigma^*_c$ given as
\begin{align}
    \overline{m}_{\Sigma_c}=\frac{2m_{\Sigma_c^*}-m_{\Sigma_c}}{3}
\end{align}
Also, using Eq. \eqref{eq:lam2corr} we can estimate $\lambda_{\Sigma,2}^b$ given in Table \ref{tab:paran2}. Now, we are well equipped to find other parameters and masses of the $n=2$ S-wave baryons. The masses of other $\Sigma$ baryons are calculated using the above computed value of $\widetilde{\overline{\Lambda}}_{\Sigma}$ and taken values of $\widetilde{\lambda}_{\Sigma,1}$ and $\widetilde{\lambda}^c_{\Sigma,2}$ in equations similar to Eq. \eqref{eq:sdcud1}. The masses of $\Sigma_c(2S)$ are shown in Table \ref{tab:Swaven2}.
The parameter $\widetilde{\overline{\Lambda}}_{\Xi'}$ can be estimated by using Eq. \eqref{eq:symm1} and the computed parameter $\widetilde{\overline{\Lambda}}_{\Sigma}$. 
The masses of $\Xi'^{(*)}_Q(2S)$ are then calculated by taking the values of $\widetilde{\lambda}_{\Xi',1}$ and $\widetilde{\lambda}^Q_{\Xi',2}$ as shown in Table \ref{tab:paran2}. These parameters are again fixed by the pattern of their corresponding parameters in Table \ref{tab:paran1}. The masses are given by equations similar to Eq. \eqref{eq:sdcud1}. These equations can be formed by taking differences of mass equations of $\Xi_c$ given by Eq. \eqref{eq:symm1} for $n=2$ and 1. Similar procedure can be employed to find the parameter $\overline{\Lambda}_{\Omega}$ and masses of $\Omega^{(*)}_{Q}(2S)$. The required parameters $\lambda_{\Omega,1}$ and $\lambda^Q_{\Omega,2}$ and shown in Table \ref{tab:paran2}, are taken similarly to their analogous parameters in Table \ref{tab:paran1}. The masses of all $\Omega(2S)$ baryons are shown in Table \ref{tab:Swaven2}. 
\begin{table}[t]
    \centering
    \begin{tabular*}{\textwidth}{@{\extracolsep{\fill}}cccc}\toprule
    \multicolumn{4}{c}{$n=2$}\\\midrule
        $\widetilde{\overline{\Lambda}}_{\Lambda}$ & $\widetilde{\lambda}_{\Lambda,1}$ & $\widetilde{\overline{\Lambda}}_{\Xi}$ & $\widetilde{\lambda}_{\Xi,1}$\\
    $1402$ & $-0.19\times10^6$ & $1574$ & $-0.21\times10^6$\\\midrule
    $\widetilde{\overline{\Lambda}}_{\Sigma}$ & $\widetilde{\lambda}_{\Sigma,1}$ & $\widetilde{\lambda}^{c}_{\Sigma,2}$ & $\widetilde{\lambda}^{b}_{\Sigma,2}$\\
    $1565$ & $-0.20\times10^6$ & $4.01\times10^4$ & $3.32\times10^4$\\\midrule
    $\widetilde{\overline{\Lambda}}_{\Xi'}$ & $\widetilde{\lambda}_{\Xi',1}$ & $\widetilde{\lambda}^{c}_{\Xi',2}$ & $\widetilde{\lambda}^{b}_{\Xi',2}$\\
    $1684$ & $-0.22\times10^6$ & $4.39\times10^4$ & $3.64\times10^4$\\\midrule
    $\widetilde{\overline{\Lambda}}_{\Omega}$ & $\widetilde{\lambda}_{\Omega,1}$ & $\widetilde{\lambda}^{c}_{\Omega,2}$  & $\widetilde{\lambda}^{b}_{\Omega,2}$\\
    1808 & $-0.21\times10^6$ & $4.83\times10^4$ & $3.64\times10^4$\\ \bottomrule
    \end{tabular*}
    \caption{The non-perturbative parameters for $n=2$ S-wave baryons. The $\widetilde{\overline{\Lambda}}$ parameters are in MeV, and parameters $\widetilde{\lambda}_1$ and $\widetilde{\lambda}_2^Q$ are in MeV$^2$.}
    \label{tab:paran2}
\end{table}
\begin{table}[t]
    \centering
    \begin{tabular*}{\textwidth}{@{\extracolsep{\fill}}ccccccccc}\toprule
     & & \multicolumn{3}{c}{$Q=c$} & \multicolumn{4 }{c}{$Q=b$}\\\cmidrule{3-5}\cmidrule{6-9}
    $J^P$ & Baryons  & Calculated & \cite{ebert2011} & \cite{PDG2021} & Calculated & \cite{ebert2011} & \cite{AKRai2022} & \cite{PDG2021} \\\midrule
    \multirow{5}{*}{$\frac{1}{2}^+$} & $\Lambda$ & \bm{$2766.6\pm2.4$} & 2769 & & $6093$ &  6089 & & $\Lambda_b(6070)$\\
                  & $\Xi$ & $2942$ & 2959 & $\Xi_c(2970)$ & $6267$ & 6266 & 6208 & \\
                  & $\Sigma$ & \textbf{2901} & 2901 & & $6246$ & 6213 & &\\
                  & $\Xi^{'}$ & $3028$ & 2983 &  & $6369$ & 6329 & 6328 &  \\
                  & $\Omega$ & $3154$ & 3088 &  & $6487$ & 6450 & 6438 &  \\ \midrule
    \multirow{3}{*}{$\frac{3}{2}^+$} & $\Sigma^*$ & $2948$ & 2936 &  & $6262$ & 6226 &  &\\
                  & $\Xi^{'*}$ & $3074$ & 3026 &  & $6381$ & 6342 & 6343 &\\
                  & $\Omega^*$ & $3190$ & 3123 &  & $6507$ & 6461 & 6462 &\\ \bottomrule
     \end{tabular*}
    \caption{The masses of 2S-wave charm and bottom baryons. All masses are in MeV.}
    \label{tab:Swaven2}
\end{table}
Using the calculated masses of $n=2$ baryons, plots are also shown in Fig. \ref{fig:ratiomb}. Again the ratio of mass terms containing $\lambda_1$ and $\overline{\Lambda}$ parameters change sign at the bottom quark mass $m_b=4620$ MeV, for both $D$ and $B$ mesons and $\Lambda_Q$ baryons. The plots look similar in both cases of $n=2$ and $n=1$. This indicates that the role of heavy quarks remains the same for the radially excited states. This feature of heavy quark mass inside the hadron highlights the effectiveness of heavy quark symmetry. For comparison, some experimentally observed excited states are  shown in Table \ref{tab:Swaven2}. We have assigned $\Lambda_b(6070)$ as 2S state. The assignment is supported by the QCD sum rule calculations of \cite{Azizi2020}. For $\Xi_c$ with $J^P=\frac{1}{2}$ the assignment is given to $\Xi_c(2970)$ as $\Xi_c(2S)$. In Ref. \cite{zezhao2020}, $\Xi_c(2970)$ is assigned as 2S wave $n_{\lambda}$ excitation. Now, note from Ref. \cite{MehenandSpringer},
\begin{align}
    \frac{m_{H^*}^{b}-m_{H}^b}{m_{H^*}^{c}-m_{H}^c}=\frac{m_{S^*}^{b}-m_{S}^b}{m_{S^*}^{c}-m_{S}^c}=\frac{m_c}{m_b}
    \label{eq:mbdividemc}
\end{align}
where, $H$ and $S$ are heavy charm and bottom meson fields with light spin $s_l=\frac{1}{2}$ and $\frac{3}{2}$ respectively. In the present paper, we have only discussed the $H$-field of mesons. A similar kind of relationship is also given by Eq. \eqref{eq:lam2corr} as $H^*$ and $H$ are hyperfine splittings of mesons which are described by $\lambda_2$ parameter. We can use \eqref{eq:mbdividemc} to study the heavy baryons and their excited states. The ratios of hyperfine splittings of charm and bottom mesons and baryons are shown in Table \ref{tab:hyperratios}. The ratios for all states can be seen to obey the Eq. \eqref{eq:mbdividemc} quite well. We can also see that the ratios remain the same for both 1S and 2S baryons. The ratio for $\Omega_Q$ and $\Omega_Q(2S)$ hyperfine are greater than the ratios of other particles. As the mass of $\Omega^*_b$ is not an experimental value but taken from Ref. \cite{Omegab32}. If we put $\frac{m_{\Omega^*_b}-m_{\Omega_b}}{m_{\Omega^*_c}-m_{\Omega_c}}=0.29$ i.e., the average of ratios for $\Sigma_Q$ and $\Xi_Q$ and compute mass of $\Omega^*_b$. We find only a 15 MeV variation in mass. The same calculation gives a 10 MeV variation for $\Omega^*_b(2S)$. Thus the ratios are very sensitive to the masses of the states and a good agreement between ratios of 1S and 2S baryons is a good assurance for calculated masses.
\begin{table}[]
    \centering
      \begin{tabular*}{\textwidth}{@{\extracolsep{\fill}}cccc}\toprule
      \multicolumn{2}{c}{$n=1$} & \multicolumn{2}{c}{$n=2$}\\\cmidrule{1-2}\cmidrule{3-4}
        States & Ratios & States & Ratios \\ \midrule
        $\frac{B^*-B}{D^*-D}$ & 0.32 & $\frac{B^*(2S)-B(2S)}{D^*(2S)-D(2S)}$ & 0.32 \vspace{0.2cm}\\ 
        $\frac{B^*_{s}-B_{s}}{D^*_s-D_s}$ & 0.34 & $\frac{B^*_{s}(2S)-B_{s}(2S)}{D^*_s(2S)-D_s(2S)}$ & 0.22 \vspace{0.2cm}\\
        $\frac{\Sigma^*_b-\Sigma_b}{\Sigma^*_c-\Sigma_c}$ & 0.30 & 
        $\frac{\Sigma^*_b(2S)-\Sigma_b(2S)}{\Sigma^*_c(2S)-\Sigma_c(2S)}$ & 0.34  \vspace{0.2cm}\\
        $\frac{\Xi^{'*}_b-\Xi^{'}_b}{\Xi^{'*}_c-\Xi^{'}_c}$ & 0.28 &
        $\frac{\Xi^{'*}_b(2S)-\Xi^{'}_b(2S)}{\Xi^{'*}_c(2S)-\Xi^{'}_c(2S)}$ & 0.26 \vspace{0.2cm} \\
        $\frac{\Omega^*_b-\Omega_b}{\Omega^*_c-\Omega_c}$ & 0.51 &
        $\frac{\Omega^*_b(2S)-\Omega_b(2S)}{\Omega^*_c(2S)-\Omega_c(2S)}$ & 0.56 \\\bottomrule
    \end{tabular*}
    \caption{Ratios of hyperfine splittings of charm and bottom meson and baryons for both ground state and radially excited $S-$wave states. For presentation purposes, we have omitted the $m$ in the superscript of states, signifying the masses of states. All masses are taken from Tables \ref{tab:Smesons}, \ref{tab:Swave} and \ref{tab:Swaven2}. }
    \label{tab:hyperratios}
\end{table}

\begin{figure}
    \centering
    \subfloat[$\Lambda_b$ and $\Xi_b$ are $s_l=0$ baryons with $J^P=\frac{1}{2}^+$]{%
    \includegraphics[width=0.32\linewidth]{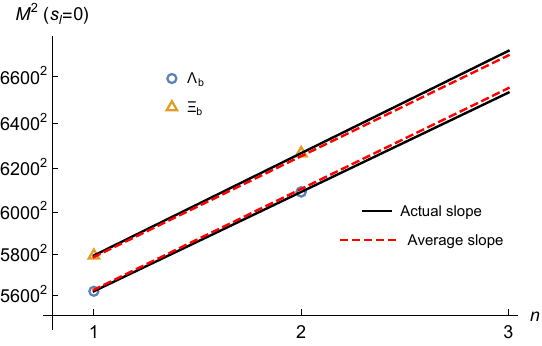}
    }\hspace{0.1cm}
    \subfloat[$\Sigma_b$, $\Xi^{'}_b$ and $\Omega_b$ are $s_l=1$ baryons with $J^P=\frac{1}{2}^+$]{%
    \includegraphics[width=0.32\linewidth]{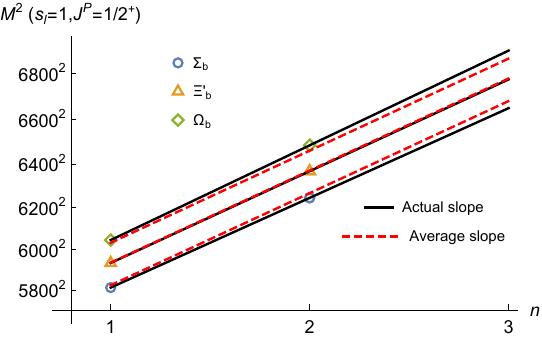}
    }\hspace{0.1cm}
    \subfloat[$\Sigma^*_b$, $\Xi^{'*}_b$ and $\Omega^*_b$ are $s_l=1$ baryons with $J^P=\frac{3}{2}^+$]{%
    \includegraphics[width=0.32\linewidth]{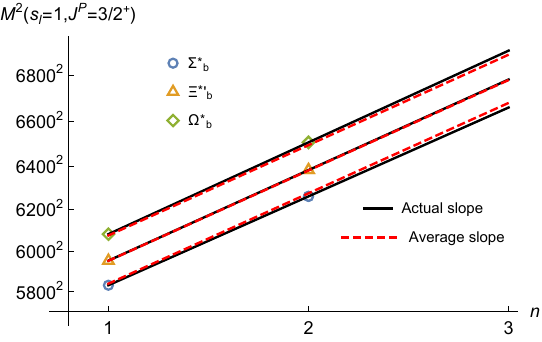}}
    \caption{Regge trajectories in ($n,M^2$) plane for bottom baryons. The solid line is the actual Regge trajectory line and the red dashed line has the average slope of all trajectories within a plot.}
    \label{fig:Regge_b}
\end{figure}

The baryons with $n=1$ and $n=2$ are used to make Regge trajectories. Regge trajectory is a very powerful method to analyze the masses of hadrons. The Regge trajectories are given by the following set of equations
\begin{align}
    J=\alpha M^2+\alpha_0\\
    n_r=\beta M^2+\beta_0
\end{align}
where, $\alpha_0$ and $\beta_0$ are intercepts and $\alpha$ and $\beta$ are slopes of the trajectories. $n_r$ is related to radial quantum number by $n_r=n-1$. $M$ and $J$ are the mass and total angular momentum of the hadrons. The masses of hadrons form a straight line in $(J,M^2)$ and $(n_r,M^2)$ planes. We have plotted the Regge trajectories in the $(n,M^2)$ plane for bottom baryons we have used and calculated in the present work. The trajectories shown in Figs. \ref{fig:Regge_b} are parallel for different baryons with the same angular quantum numbers. Also, the different trajectories within the same plot are equally spaced. These properties of Regge trajectories signify a good set of masses of hadrons. The non-perturbative parameters of HQET are a very useful method to identify the masses of excited states. Confirming these radially excited baryons will help further fix the parameters given in Table \ref{tab:paran2}. 
\section{Decays and couplings}
\subsection{Strong decays}
\begin{figure}
    \centering
    \includegraphics[width=10cm]{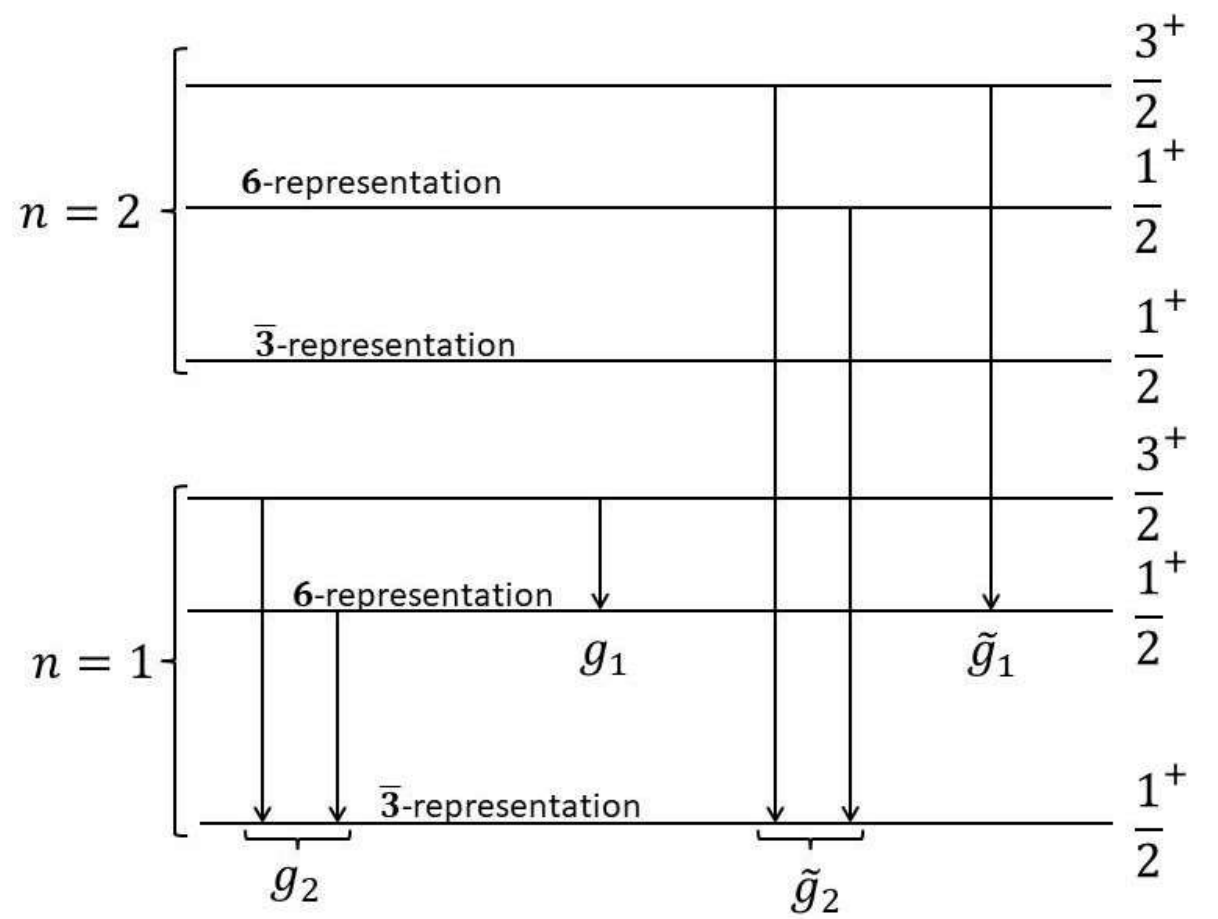}
    \caption{Transitions of baryon decay with their coupling constants.}
    \label{fig:transition}
\end{figure}

The charm and bottom hadrons provide a unique platform to study the non-perturbative nature of QCD, as the charm and bottoms are heavier than the QCD scale. We aim to study the strong decays of both 1S and 2S charm and bottom baryons with a pseudoscalar meson as an outgoing particle with baryon. The isospin conserving decays are considered. The S-wave charm and bottom baryons decaying to S-wave baryons contain the P-wave couplings $g_1$ and $g_2$, which are shown in Fig. \ref{fig:transition}. The Lagrangian \eqref{eq:lag} with baryon fields \eqref{eq:tfield} and \eqref{eq:sfield} given in Ref. \cite{pirjol1997} is used to compute the strong decay widths of charm and bottom baryons. The strong decay widths with P-wave couplings for S-wave heavy baryons emitting a pseudoscalar meson are given as \cite{pirjol1997}
\begin{align}
    \Gamma\left(\Sigma^{(*)}_Q\rightarrow\Lambda_Q \pi\right)&=\frac{g_2^2}{2\pi f_{\pi}^2}\frac{m(\Lambda_Q)}{m(\Sigma^{(*)}_Q)}\abs{p_{\pi}}^3 \label{eq:g1232}\\
    \Gamma\left(\Sigma_Q^*\rightarrow\Sigma_Q \pi\right)&=\frac{g_2^2}{16\pi f_{\pi}^2}\frac{m(\Sigma_Q)}{m(\Sigma^*_Q)}\abs{p_{\pi}}^3 \label{eq:g32}
\end{align}
where, $g_1$ and $g_2$ are coupling constants and $p_{\pi}$ is the emitted pseudoscalar meson momentum. The Eq. \eqref{eq:g1232} gives decays for baryons from $\bf{6}$ to $\bf{\bar{3}}$ representation and Eq. \eqref{eq:g32} gives decays from $\bf{6}$ to $\bf{6}$ representation baryons. The decay widths for $1S\rightarrow1S$ are shown in Table \ref{tab:strong1s1s}. The decay widths of $\Sigma_c$ and $\Sigma_c^*$ are given in terms of coupling $g_2^2$ in Table \ref{tab:strong1s1s}. On comparing decay widths with experimental values, we get an average value of coupling $\abs{g_2}=0.559^{+0.006}_{-0.010}$. The strong transition $\Sigma_c^*\rightarrow\Sigma_c\pi$ is kinematically forbidden. Thus, coupling $g_1$ can not be estimated directly through their widths. The coupling $g_1$ is then estimated using quark model relation between $g_1$ and $g_2$ \cite{pirjol1997},
\begin{align}
\abs{g_1}=2\sqrt{\frac{2}{3}}\abs{g_2}=0.913^{+0.010}_{-0.017} \label{eq:g1g2}
\end{align}
Our calculated couplings $g_1$ and $g_2$ are comparable to the estimated couplings in Ref. \cite{pirjol1997,TMYan1992}. In Table \ref{tab:strong1s1s}, the widths of all $1S\rightarrow1S$ transitions are given for other baryons using the above values of the $g_1$ and $g_2$ couplings. The experimental values of decay widths are also given with the corresponding states in Table \ref{tab:strong1s1s}, which are in good agreement with calculated widths. 

For decays of $n=2$ S-wave baryons, the couplings $g_1$ and $g_2$ are changed to $\widetilde{g}_1$ and $\widetilde{g}_2$. The Eqs. \eqref{eq:g1232} and \eqref{eq:g32} are used to calculate the transitions of $2S$ and $1S$ baryons with the corresponding couplings are shown in Fig. \ref{fig:transition}. There may be other transitions possible for 2S baryons, so we will not calculate the total decay widths of the particles. The third column of Table \ref{tab:strong2S1S} contains the calculated partial decay widths in terms of couplings.
\begin{table}
    \centering
    \begin{tabular*}{\textwidth}{@{\extracolsep{\fill}}cccc}\toprule
    State & Mode & Calculated(MeV) &  \cite{PDG2021} (MeV)  \\\midrule
     $\Sigma^{++}_c$ & $\Lambda^+_c \pi^+$ & $(6.26\pm0.07) g_2^2$ & $1.89_{-0.18}^{+0.09}$\\
     $\Sigma^{0}_c$ & $\Lambda^+_c \pi^-$ & $(6.18\pm0.07) g_2^2$ & $1.83_{-0.19}^{+0.11}$\\
     $\Sigma_c^{*++}$ & $\Lambda^+_c \pi^+$ & $\left(46.03^{+0.20}_{-0.18}\right) g_2^2$ & $14.78_{-0.40}^{+0.30}$\\
     $\Sigma_c^{*0}$ & $\Lambda^+_c \pi^-$ & $(46.10\pm0.19) g_2^2$ & $15.3_{-0.5}^{+0.4}$ \\ \midrule
     $\Xi_{c}^{'*+}$ & $\Xi^0_c \pi^+$ & $2.78^{+0.08}_{-0.11}$ & $2.14\pm0.19$\\ 
     $\Xi_c^{'*0}$ & $\Xi^+_c \pi^-$ & $3.33_{-0.15}^{0.12}$ & $2.35\pm0.18\pm0.13$\\ \midrule
      $\Sigma_b^+$ & $\Lambda_b^0 \pi^+$ & $5.90^{+0.13}_{-0.22}$ & $5.0\pm0.5$\\
      $\Sigma_b^-$ & $\Lambda_b^0 \pi^-$ & $6.93^{+0.16}_{-0.26}$ & $5.3\pm0.5$\\
     $\Sigma_b^{*+}$ & $\Lambda_b^0 \pi^+$ & $10.35^{+0.23}_{-0.40}$ & $9.4\pm0.5$\\
     $\Sigma_b^{*-}$ & $\Lambda_b^0 \pi^-$ & $11.51^{+0.25}_{-0.40}$ & $10.4\pm0.8$\\
     $\Xi_b^{'*0}$ & $\Xi_b^- \pi^+$ & $0.89^{+0.07}_{-0.08}$ & $0.90\pm0.16\pm0.08$\\ 
     $\Xi_{b}^{'*-}$ & $\Xi_b^0 \pi^-$ & $1.69^{+0.06}_{-0.08}$ & $1.65\pm0.31\pm0.10$\\ \bottomrule
    \end{tabular*}
    \caption{Decays of 1S charm and bottom baryons. The first four decays of $\Sigma_c$ and $\Sigma_c^*$ are used to compute, $\abs{g_2}=0.559^{+0.006}_{-0.010}$. All other decays are computed using the above coupling values.}
    \label{tab:strong1s1s}
\end{table}
\begin{table}[]
    \centering
    \begin{tabular*}{\textwidth}{@{\extracolsep{\fill}}ccc}\toprule
      State(2S) & Mode & Present (MeV)\\\midrule
    $\Sigma_c$ & $\Lambda_c \pi$ & 1102.53$\widetilde{g}_2^2$ \\ \midrule
    $\Sigma_c^*$ & $\Lambda_c \pi$ & 1339.82 $\widetilde{g}_2^2$\\ 
                & $\Sigma_c \pi$ & 77.9941 $\widetilde{g}_1^2$ \\ \midrule 
    $\Xi_c^{'}$ & $\Lambda_c K$ & $779.447 \widetilde{g}_2^2$ \\ 
                & $\Xi_c \pi$ & 882.941 $\widetilde{g}_2^2$ \\ \midrule 
    $\Xi_c^{'*}$ & $\Lambda_c K$ & $1027.16 \widetilde{g}_2^2$ \\ 
                & $\Xi_c \pi$ & 1096.23 $\widetilde{g}_2^2$ \\ 
                & $\Sigma_c K$ & 34.7414 $\widetilde{g}^2_1$ \\
                & $\Xi_c^{'} \pi$ & 80.026 $\widetilde{g}_1^2$ \\ \midrule
    $\Omega_c^{*}$ & $\Xi_c K$ & 707.294 $\widetilde{g}_2^2$ \\ 
                & $\Omega_c \pi$ & 81.0173 $\widetilde{g}_1^2$ \\ 
                & $\Xi_c^{'} K$ & 31.7231$\widetilde{g}_1^2$ \\ \midrule\midrule 
    $\Sigma_b$ & $\Lambda_b \pi$ & 1604.43 $\widetilde{g}_2^2$ \\\midrule  
    $\Sigma_b^*$ & $\Lambda_b \pi$ & 1726.44 $\widetilde{g}_2^2$ \\ 
                & $\Sigma_b \pi$ & 73.862 $\widetilde{g}_1^2$ \\ \midrule 
    $\Xi_b^{'}$ & $\Xi_b \pi$ & 1253.4 $\widetilde{g}_2^2$ \\ 
                & $\Lambda_b K$ & 1201.17 $\widetilde{g}_2^2$ \\ \midrule 
    $\Xi_b^{'*}$ & $\Lambda_b K$ & 1299.49 $\widetilde{g}_2^2$ \\ 
                & $\Xi_b \pi$ & 1332.52 $\widetilde{g}_2^2$ \\ 
                & $\Sigma_b K$ & 20.0198 $\widetilde{g}^1_2$ \\ 
                & $\Xi_b^{'} \pi$ & 72.6214 $\widetilde{g}_1^2$ \\ \midrule 
    $\Omega_b^*$ & $\Omega_b \pi$ & 80.7825 $\widetilde{g}_1^2$  \\ 
                & $\Xi_b K$ & 929.655$\widetilde{g}_2^2$ \\ 
                & $\Xi_b^{'} K$ & 21.8842 $\widetilde{g}_1^2$ \\\bottomrule 
    \end{tabular*}
    \caption{The strong decay widths of 2S-wave charm baryons in terms of couplings $g_1$ and $g_2$.}
    \label{tab:strong2S1S}
\end{table}
We have computed the widths of $\Sigma_c(2S)\rightarrow \Lambda_c \pi$, the other modes $\Sigma_c\pi$ and $\Sigma_c^*\pi$ are also possible, which are not calculated here. So, we have used the partial widths to compare with other models and calculate the ratio of couplings $\frac{\widetilde{g}_1}{\widetilde{g}_2}$. For $\Xi^{'}_c(2S)$, our estimation show an enhancement in $\Gamma(\Xi^{'}_c(2S)\rightarrow\Xi_c\pi)$ channel when compared to $\Gamma(\Xi^{'}_c(2S)\rightarrow\Lambda_c K$ channel. Whereas using $^{3}P_0$ model Ze Zhao \cite{zezhao2020} calculated decay widths of 2S strange baryons to give a assignment to $\Xi_{c}(2970)$ as a $2S$ radial excitation found an enhancement in $\Gamma(\Xi^{'}_c(2S)\rightarrow\Lambda_c K$ channel with respect to $\Gamma(\Xi^{'}_c(2S)\rightarrow\Xi_c\pi$ channel. Although the values of ratios of these channels are close to unity as shown in Eq \eqref{eq:Xidecayratio}. The pattern is similar for $\Xi_c^{'*}$ and for the bottom sector also. 
\begin{subequations}
\begin{align}
    \mathcal{R}_1= \frac{\Gamma(\Xi^{'}_c(2S)\rightarrow\Xi_c\pi)}{\Gamma(\Xi^{'}_c(2S)\rightarrow\Lambda_c K)} = 1.13 = 0.90~ \cite{zezhao2020}  \approx 1 \\  
    \mathcal{R}_2=\frac{\Gamma(\Xi^{'*}_c(2S)\rightarrow\Xi_c\pi)}{\Gamma(\Xi^{'*}_c(2S)\rightarrow\Lambda_c K)} = 1.07 = 0.90~ \cite{zezhao2020} \approx 1 \\
    \mathcal{R}_3=\frac{\Gamma(\Xi^{'}_b(2S)\rightarrow\Xi_b\pi)}{\Gamma(\Xi^{'}_b(2S)\rightarrow\Lambda_b K)} = 1.04 = 0.94~ \cite{HuiZenHe2021} \approx 1 \\
    \mathcal{R}_4=\frac{\Gamma(\Xi^{'*}_b(2S)\rightarrow\Xi_b\pi)}{\Gamma(\Xi^{'*}_b(2S)\rightarrow\Lambda_b K)} = 1.03 = 0.94~ \cite{HuiZenHe2021} \approx 1 
    \end{align}
    \label{eq:Xidecayratio}
\end{subequations}
So using Eqs. \eqref{eq:Xidecayratio}, both $\Xi_Q \pi$ and $\Lambda_Q \pi$ channels can't be used to differentiate between $\Xi_Q^{'}$ and $\Xi_Q^{'*}$ states. The total decay widths for $\Xi_c^{'}$ and $\Xi_c^{'*}$ calculated by Ze Zhao \cite{zezhao2020} are 356.1 MeV and 311.4 MeV (for $n_{\rho}$ excited state) and, 53.0 MeV and 40.9 MeV (for $n_{\lambda}$ excited state), respectively. Using the non-relativistic constituent quark model \cite{Chen2017}, the charm baryons are studied, and total decay widths are calculated for $\Xi_c^{'}$ and $\Xi_)c^{'*}$ to be 55.47 MeV and 47.91 MeV respectively. In Ref. \cite{HuiZenHe2021} bottom strange baryons are studied using $^{3}P_0$ model. They have predicted the total decay for $\Xi_b^{'}$ and $\Xi_b^{'*}$ to be 33.89 MeV and 35.64 MeV. To get an estimation of the couplings, the following  ratios of partial widths may be used, 
\begin{subequations}
    \begin{align}
        \mathcal{R}_5&=\frac{\Gamma(\Xi^{'*}_c(2S)\rightarrow\Sigma_c K)}{\Gamma(\Xi^{'*}_c(2S)\rightarrow\Lambda_c K)} = 0.034 \frac{\widetilde{g}_1^2}{\widetilde{g}_2^2} = 0.010~ \cite{zezhao2020} = 0.35~ \cite{Chen2017}  \\
        \mathcal{R}_6&=\frac{\Gamma(\Xi^{'*}_c(2S)\rightarrow\Xi^{'}_c \pi)}{\Gamma(\Xi^{'*}_c(2S)\rightarrow\Lambda_c K)} = 0.078 \frac{\widetilde{g}_1^2}{\widetilde{g}_2^2} = 0.133 ~ \cite{zezhao2020} = 0.191~ \cite{Chen2017}  \\
        \mathcal{R}_7&=\frac{\Gamma(\Xi^{'*}_b(2S)\rightarrow\Sigma_b K)}{\Gamma(\Xi^{'*}_b(2S)\rightarrow\Lambda_b K)}= 0.015 \frac{\widetilde{g}_1^2}{\widetilde{g}_2^2} = 0.078 ~\cite{HuiZenHe2021}  \\
        \mathcal{R}_8&=\frac{\Gamma(\Xi^{'*}_b(2S)\rightarrow\Xi^{'}_b \pi)}{\Gamma(\Xi^{'*}_b(2S)\rightarrow\Lambda_b K)}= 0.056 \frac{\widetilde{g}_1^2}{\widetilde{g}_2^2} = 0.146 ~\cite{HuiZenHe2021} 
    \end{align}
    \label{eq:xidecayratio2}
\end{subequations}
The Eqs. \eqref{eq:xidecayratio2} gives a discrepancy in the prediction for ratio $\mathcal{R}_5$ by Ref. \cite{zezhao2020} and \cite{Chen2017}, so we are not using it to draw any results. We have used both values of $\mathcal{R}_6$ from Refs. \cite{zezhao2020} and \cite{Chen2017} to find average ratio of couplings $\frac{\widetilde{g}_1}{\widetilde{g}_2}=1.43$. From $\mathcal{R}_8$, the ratio of couplings $\frac{\widetilde{g_1}}{\widetilde{g}_2}$ is 1.61. The average of both values give $\frac{\widetilde{g}_1}{\widetilde{g}_2}=1.52$. Similar to ratio $\mathcal{R}_5$ for $\Xi^{'*}_c$; we are not considering $\mathcal{R}_7$ for the bottom sector also. The above ratio of couplings is considerably close to the quark model relation given in Eq. \eqref{eq:g1g2} for couplings $g_1$ and $g_2$ of the ground state ($n=1$). The partial decay width ratios $\mathcal{R}_5$, $\mathcal{R}_6$, $\mathcal{R}_7$ and $\mathcal{R}_8$ shows that in non-relativistic constituent quark model \cite{Chen2017}, $^{3}P_0$ model \cite{zezhao2020} and in the present analysis, the $\Lambda_Q K$ channel is dominant decay mode for $\Xi_Q^{'*}$ in both charm and bottom sectors. The experimental facilities may provide information in the future to confirm the states, their decay widths, and couplings.
\subsection{Semi-electronic decay rates}
The heavy quark symmetry gives many ways to study heavy-light hadronic systems. The hyperfine mass splittings, non-perturbative parameters, and strong coupling constants are some of the properties which do not change with heavy quark flavor. These symmetries allow the studies to go from charm to bottom states. As charm hadrons are widely explored in comparison to bottom hadrons, heavy quark symmetry is very useful for predicting the properties of states containing bottom quarks. The strong widths are already discussed in the previous section 5.1. We move forward by including the heavy flavor conserving semi-electronic decay rates of $n=1$ and $n=2$ baryons. In these transitions, the heavy quark is considered a spectator during the decay transition, and the strange quark decays into an up quark with the help of W-meson. The assumption of heavy quark as a spectator may give a way to study the nature of heavy quark in the radial excitations. The semi-electronic decay formulae for singly heavy baryons are given below\cite{Faller2015},
\begin{align}
    \Gamma^{1/2^+\rightarrow1/2^+}_{0^+\rightarrow0^+}&=\frac{G_F^2|V_{CKM}|^2}{60 \pi^3}(\Delta m)\label{eq:semiel1}\\
    \Gamma^{1/2^+\rightarrow1/2^+}_{0^+\rightarrow1^+}&=\frac{G_F^2|V_{CKM}|^2}{60 \pi^3}(\Delta m)   \label{eq:semiel2}\\ \Gamma^{1/2^+\rightarrow3/2^+}_{0^+\rightarrow1^+}&=2\Gamma^{1/2^+\rightarrow1/2^+}_{0^+\rightarrow1^+}\label{eq:semiel3}\\
    \Gamma^{1/2^+\rightarrow1/2^+}_{1^+\rightarrow1^+}&=\frac{G_F^2|V_{CKM}|^2}{15 \pi^3}(\Delta m)
    \label{eq:semiel4}
    \end{align}
where, the superscripts are $J^P$ and subscripts are light degrees of freedom ($s_{l_i}\rightarrow s_{l_f}$) of initial and final baryons. $\Delta m$ is the mass difference between the final and initial baryons. The Eq. \eqref{eq:semiel1} gives the decay rate of $\Xi\rightarrow\Lambda e \nu$. The Eqs. \eqref{eq:semiel2} and \eqref{eq:semiel3} are for decays having final states with a "$\Sigma$-like" baryon. The Eq. \eqref{eq:semiel4} gives decay widths for a "$\Xi$-like" baryon in the final state. The Fermi coupling constant $G_F=1.166 \times 10^{-5}$ GeV$^-2$ and Cabibbo-Kobayashi-Maskawa (CKM) matrix element $V_{CKM}=V_{ud}=0.224$ are taken from PDG\cite{PDG2021}.  The masses of baryons are used from Table \ref{tab:Swave} and \ref{tab:Swaven2}. The calculated decay rate for transitions: $n=1\rightarrow n=1$, $n=2\rightarrow n=2$ and $n=2\rightarrow n=1$, are tabulated in the table below.
\begin{table}[htb]
    \centering
    \begin{tabular*}{\textwidth}{@{\extracolsep{\fill}}ccccc}\toprule
    Mode  & $J_i^{P_i}\rightarrow J_f^{P_f}$ & $s_{l_i}\rightarrow s_{l_f}$ & Decay Rate & Ref \cite{Faller2015}\\ \midrule
    $\Xi_c\rightarrow\Lambda_c e \nu$ & $\frac{1}{2}^+\rightarrow \frac{1}{2}^+$ &  $0\rightarrow0$  & $7.49 \times 10^{-19}$ & $7.91\times10^{-19}$\\
    $\Xi_c\rightarrow\Sigma_c e \nu$ & $\frac{1}{2}^+\rightarrow\frac{1}{2}^+$ & $0\rightarrow1$ & $3.34\times10^{-24}$ & $3.74\times10^{-24}$\\
    $\Omega_c\rightarrow\Xi_c e \nu$  & $\frac{1}{2}^+\rightarrow\frac{1}{2}^+$ & $1\rightarrow0$ & $2.18\times10^{-18}$ & $2.26\times 10^{-18}$\\
    $\Omega_c\rightarrow\Xi'_c e \nu$ & $\frac{1}{2}^+\rightarrow\frac{1}{2}^+$ & $1\rightarrow1$ & $3.19 \times 10^{-19}$ & $3.63\times10^{-19}$ \\\bottomrule
    \end{tabular*}
    \caption{Semi-electronic decays of ($1S\rightarrow 1S$) charm baryons.}
    \label{tab:n1n1c}
\end{table}
\begin{table}[htb]
    \centering
    \begin{tabular*}{\textwidth}{@{\extracolsep{\fill}}cccc}\toprule
    Mode  & $J_i^{P_i}\rightarrow J_{f}^{P_f}$ & $s_{l_i}\rightarrow s_{l_f}$ & Decay Rate \\ \midrule
    $\Xi_c\rightarrow\Lambda_c e \nu$ & $\frac{1}{2}^+\rightarrow \frac{1}{2}^+$ &  $0\rightarrow0$  & $6.12 \times 10^{-19}$\\
    $\Xi_c\rightarrow\Sigma_c e \nu$ & $\frac{1}{2}^+\rightarrow\frac{1}{2}^+$ & $0\rightarrow1$ & $4.27\times10^{-22}$\\
    $\Omega_c\rightarrow\Xi_c e \nu$  & $\frac{1}{2}^+\rightarrow\frac{1}{2}^+$ & $1\rightarrow0$ & $1.58\times10^{-18}$\\
    $\Omega_c\rightarrow\Xi'_c e \nu$ & $\frac{1}{2}^+\rightarrow\frac{1}{2}^+$ & $1\rightarrow1$ & $4.68 \times 10^{-19}$\\\bottomrule
    \end{tabular*}
    \caption{Semi-electronic decays of ($2S\rightarrow 1S$) charm baryons}
    \label{tab:n2n1c}
\end{table}
\begin{table}[htb]
    \centering
    \begin{tabular*}{\textwidth}{@{\extracolsep{\fill}}cccc}\toprule
    Mode  & $J_i^{P_i}\rightarrow J_f^{P_f}$ & $s_{l_i}\rightarrow s_{l_f}$ & Decay Rate \\ \midrule
    $\Xi_c\rightarrow\Lambda_c e \nu$ & $\frac{1}{2}^+\rightarrow \frac{1}{2}^+$ &  $0\rightarrow0$  & $4.46 \times 10^{-16}$\\
    $\Xi_c\rightarrow\Sigma_c e \nu$ & $\frac{1}{2}^+\rightarrow\frac{1}{2}^+$ & $0\rightarrow1$ & $1.03\times10^{-16}$\\
    $\Omega_c\rightarrow\Xi_c e \nu$  & $\frac{1}{2}^+\rightarrow\frac{1}{2}^+$ & $1\rightarrow0$ & $5.56\times10^{-16}$\\
    $\Omega_c\rightarrow\Xi'_c e \nu$ & $\frac{1}{2}^+\rightarrow\frac{1}{2}^+$ & $1\rightarrow1$ & $9.31 \times 10^{-16}$\\\bottomrule
    \end{tabular*}
    \caption{Semi-electronic decays of ($2S\rightarrow 2S$) charm baryons}
    \label{tab:n2n2c}
\end{table}
The transition rates for $1S\rightarrow 1S$ are in good agreement with Ref. \cite{Faller2015}. The rates for $2S \rightarrow 1S$ decays are very similar to the $1S\rightarrow 1S$ rates indicating that the heavy quark remains mostly unaffected by the presence of light quarks in both 1S and 2S states. This support the dependence of higher mass terms on $\frac{1}{m_Q}$. As the binding energy for radial excitations is suppressed by $\frac{1}{m_Q}$ factor, the heavy quark plays a dominant role. Thus, being a spectator it does not affect the decay rate of $1S \rightarrow 1S$. The decay rates of $2S \rightarrow 2S$ transitions are higher than the other transitions. This enhancement may be due to transitions between the same radial levels and a small mass difference between 2S hadrons. This makes the available phase space for transitions large which results in higher decay rates. In  $2S\rightarrow2S$ decays, the $\Omega_c(2S)$ has higher rate than $\Xi_c(2S)$, while in both $1S\rightarrow1S$ and $2S\rightarrow1S$ transitions, $\Xi_c\rightarrow\Lambda e \nu$ has greater decay rate than $\Omega_c$.
\section{Conclusion}
The radially excited heavy baryons are analyzed in the framework of HQET. The non-perturbative parameters of HQET are calculated for $n=1$ heavy baryons. The bottom quark mass in the charm and bottom mesons and baryons is varied to analyze the behavior of HQET parameters $\overline{\Lambda}$, $\lambda_1$ in the mass terms of HQET mass formulae. The non-perturbative parameter $\overline{\Lambda}$ are calculated for $n=2$ baryons, using the HQET symmetry of the parameters. The masses of $n=2$ heavy baryons are estimated by the help of these non-perturbative parameters. The variation of masses of baryons on parameters $\lambda_1$ and $\lambda_2$ are also shown. The calculated masses are compared with other theoretical and experimental results. The Regge trajectories of $n=1$ and $n=2$ S-wave baryons are also shown for the masses obtained. The Regge trajectories are parallel and equidistant for the same quantum numbers ($J^P$). The strong decays of the ground state ($n=1$) and radially excited states ($n=2$) are studied, and couplings $g_1$ and $g_2$ are estimated. The ratios of partial decay widths of 2S charm and bottom baryons are compared with other theoretical models. The ratios $\mathcal{R}_1 - \mathcal{R}_4$ shows that the modes $\Lambda_Q K$ and $\Xi_Q \pi$ may not be helpful to distinguish the $\Xi^{'}_Q(2S)$ with $J^P=\frac{1}{2}^+$ and $\Xi^{'*}_Q(2S)$ with $J^P=\frac{3}{2}^+$ states. The ratios $\mathcal{R}_5 - \mathcal{R}_8$ shows an enhancement in width for $\Lambda_Q K$ mode for $\Xi^{'*}_Q(2S)$ in respect to $\Sigma_Q K$ mode. So, these channels may be used to identify the $\Xi^{'*}_Q(2S)$ state with $J^P=\frac{3}{2}^+$. Also, the ratio of couplings is estimated using the ratios of partial widths of $\Xi^{'*}_Q(2S)$ baryons to be $\frac{\widetilde{g}_1}{\widetilde{g}_2}=1.52$. The calculated value is very close to the ratio of couplings given by the quark model for $g_1$ and $g_2$ mentioned in Eq. \eqref{eq:g1g2} for $n=1$ case. The semi-electronic decays in the spectator heavy quark assumption are studied for charm baryons for transitions $1S\rightarrow 1S$, $2S\rightarrow 1S$, and $2S\rightarrow 2S$. The decay rates for $1S\rightarrow 1S$ and $2S\rightarrow 1S$ are of the same order. These decays show the dominating role played by the heavy quark inside the hadrons. The present study may be helpful in near future to confirm the radial excitation of singly heavy baryons. 
\section{Acknowledgment}
The authors gratefully acknowledge the financial support by the
Department of Science and Technology \\(SERB/F/9119/2020), New
Delhi and for Junior Research Fellowship (09/0677(11306)/2021-EMR-I) by Council of Scientific and Industrial Research, New Delhi.
\bibliographystyle{epj}
\bibliography{reference}

\end{document}